\documentclass{aa}

\usepackage{natbib}
\bibpunct{(}{)}{;}{a}{}{,} 

\usepackage{txfonts}
\usepackage{siunitx}
\usepackage{xurl}
\usepackage{xcolor}
\usepackage{float}
\usepackage{upgreek}

\usepackage{graphicx}   
\usepackage{bm}
\usepackage{amsmath}

\newcommand{\rmn}[1]{\mathrm{#1}}

\usepackage{xpatch}
\usepackage{hyperref}
\hypersetup{
	colorlinks=true,
	breaklinks=true,
	citecolor=blue,
	allcolors=blue,
	frenchlinks=true
}

\defcitealias{Ehlert2023}{E23}
\defcitealias{Jlassi2026a}{J26}

\makeatletter
\xpatchcmd\NAT@citex
{
	\@citea\NAT@hyper@{
		\NAT@nmfmt{\NAT@nm}
		\hyper@natlinkbreak{\NAT@aysep\NAT@spacechar}{\@citeb\@extra@b@citeb}
		\NAT@date
	}
}
{
	\@citea
	\NAT@nmfmt{\NAT@nm}
	\NAT@aysep\NAT@spacechar
	\NAT@hyper@{\NAT@date}
}
{}{}
\xpatchcmd\NAT@citex
{
	\@citea\NAT@hyper@{
		\NAT@nmfmt{\NAT@nm}
		\hyper@natlinkbreak{\NAT@spacechar\NAT@@open\if*#1*\else#1\NAT@spacechar\fi}
		{\@citeb\@extra@b@citeb}
		\NAT@date
	}
}
{
	\@citea
	\NAT@nmfmt{\NAT@nm}
	\NAT@spacechar\NAT@@open\if*#1*\else#1\NAT@spacechar\fi
	\NAT@hyper@{\NAT@date}
}
{}{}
\makeatother

\makeatletter
\renewcommand*\aa@pageof{, page \thepage{} of \pageref*{LastPage}}
\makeatother

\graphicspath{/figures/}

\begin{document}

\title{Simulating realistic radio morphologies of Fanaroff-Riley I jets in a self-regulating  cool-core cluster}
 \titlerunning{Simulating FRI jets in a self-regulating cool-core cluster}
    
\author{L\'{e}na Jlassi\inst{1,2}
	\and
	Rainer Weinberger\inst{1}
	\and
	Christoph Pfrommer\inst{1}
	\and
	Maria Werhahn\inst{3}
	\and
	Joseph Whittingham\inst{1}
	\and
	Philipp Girichidis\inst{4}
}

\institute{Leibniz-Institut f\"{u}r Astrophysik Potsdam (AIP), 
	An der Sternwarte 16, D-14482 Potsdam, Germany\\
	\email{ljlassi@aip.de}
	\and
	Institut für Physik und Astronomie, Universität Potsdam, Karl-Liebknecht-Str. 24/25, 14476 Potsdam, Germany
	\and
	Max-Planck-Institut f\"{u}r Astrophysik, Karl-Schwarzschild-Str. 1, 85748 Garching, Germany
	\and
	Universit\"{a}t Heidelberg, Zentrum f\"{u}r Astronomie, Institut f\"{u}r Theoretische Astrophysik, Albert-Ueberle-Str. 2, 69120 Heidelberg, Germany \label{ITA}
}

\date{\today}

\abstract{
	
Active galactic nucleus (AGN) jets radiate radio synchrotron emission displaying a wide range of morphologies. At the same time, they provide heat to prevent cooling flows in cool-core galaxy clusters. We produce mock radio observations of AGN jets in a self-regulating cool-core galaxy cluster. To this end, we employ magneto-hydrodynamical simulations of an idealised Perseus-like galaxy cluster, in which accretion-powered low-density jets accelerate cosmic ray protons and electrons by means of a sub-grid model. Cosmic ray electron spectra are spatially and temporally evolved along Lagrangian tracer trajectories using the Fokker-Planck solver \textsc{Crest} to produce radio synchrotron emission. Self-regulated AGN jets stabilize the cool-core cluster against cooling flows and produce realistic Fanaroff-Riley I (FRI) and disturbed lobe morphologies, in contrast to symmetrical lobe structures obtained with a single jet outburst of fixed power. Our mock radio observations are viewed in a blazar configuration -- along the jet axis -- and exhibit complex radio-emitting lobe structures despite this. This highlights the strong deflection of light jets by cold gas structures and suggests that small-scale black hole and jet properties cannot be inferred from kpc-scale FRI radio lobe morphologies. Combining self-consistently evolved magnetic fields and electron spectra enables us to explain a known observational phenomenon, whereby radio observations of AGN lobes on galaxy cluster scales occasionally display similar spatial extents at different frequencies: in 1--50~$\mu \rmn{G}$ magnetic fields obtained in our cool-core environment, both freshly accelerated and hundreds-of-Myr-old electrons are able to contribute to the 150~MHz--1.4~GHz frequency range. In addition to providing a window into AGN jet feedback through radio diagnostics, these simulations employing a time-varying jet luminosity yield radio phenomena typical of episodic AGN activity, such as restarted and remnant radio galaxies.
}


\maketitle
\nolinenumbers

\section{Introduction}

The cooling time of the X-ray emitting gas in a large number of galaxy clusters is less than the Hubble time \citep{Peterson2006}. In these so-called `cool-core' galaxy clusters, the expected cooling flows are not observed and instead, the total star formation rates are one to two orders of magnitude below the predicted mass deposition rates \citep{Crawford1999}. Jets launched in the central active galactic nucleus (AGN) system constitute a mechanism to heat the intra-cluster medium \citep[ICM;][]{Fabian2012, McNamara2012, Hlavacek-Larrondo2022}. The jets' mechanical luminosities estimated using X-ray cavity powers appears well correlated with the X-ray cooling luminosity \citep{Birzan2004, Rafferty2006, Kokotanekov2017}, emphasizing the self-regulating nature of these systems.

Radio lobes inflated by the jets are often observed at the location of these X-ray cavities \citep{McNamara2000, Birzan2004, Birzan2008} and are hence complementary tracers of the energy released by the central supermassive black hole (SMBH). This radio synchrotron emitting plasma indicates the presence of relativistic particles, so-called cosmic rays (CRs), and magnetic fields. These CRs are likely accelerated through diffusive shock acceleration \citep{Axford1977, Bell1978, Bell1978a} and Fermi second order processes \citep{Fermi1949}, although acceleration processes specific to jet dynamics are diverse \citep[see e.g.][for a review]{Matthews2020}.

The total pressure measured in AGN jet lobes points to the presence of a non-radiating component, likely in the form of CR protons \citep{Dunn2004, Birzan2008, Croston2008, Croston2018}. This hadronic component can also provide heating to the ICM by excitation of Alfv\'en waves through plasma instabilities and collisionless damping of these waves \citep{Guo2008, Pfrommer2013, Jacob2017a, Jacob2017}. For this reason CR protons have been included in numerous studies simulating AGN feedback in cool-core clusters \citep{Sijacki2008, Guo2011, Ruszkowski2017, Ehlert2018, Wang2020, Su2021, Beckmann2022}.

Through synchrotron radiation, CR electrons provide a direct observational window into the dynamics of AGN jets in such environments. The Fanaroff-Riley (FR) morphological classification \citep{Fanaroff1974} has been used to identify core-brightened radio-emitting jets -- referred to as radio galaxies -- as Fanaroff-Riley Type I (FRI) and edge-brightened radio galaxies as Fanaroff-Riley Type II (FRII) \citep[see][for a review]{Hardcastle2020}. This classification is also linked with the source's radio luminosity, with FRI radio galaxies having lower luminosities on average in comparison to FRIIs \citep{An2012, Mingo2019}. The phenomenology underlying this morphological divide is generally described as FRIIs remaining relativistic and collimated on large scales, often terminating in hotspots, while FRIs, although initially relativistic, decelerate on kiloparsec scales upon interacting with the environment and by entraining the surrounding plasma \citep{Bicknell1995, Laing2002}. The environment hosting these sources has hence been inferred to explain the classification, with richer and more massive environments such as those of galaxy clusters where FRI radio galaxies are predominant \citep{Wing2011, Mingo2019}, providing an explanation for the halting of jets on small scales \citep{Ledlow1996}. Nonetheless, host environment alone has proved insufficient to explain the divide \citep{Best2007, Capetti2017a, Mahatma2023}. Instead, it is currently thought to be caused by the interplay of intrinsic black hole and jet properties \citep{Lin2010, Best2012, Ineson2015, Tadhunter2016} with the host environmental density \citep{Bicknell1995, Kaiser2007, Croston2008}.

Furthermore, there exists a variety of extended sources with characteristics deviating from the twin-lobed jets included in the bimodal FR classification. Classifications associated with the episodic nature of AGNs include remnant radio galaxies which describe fading AGN accelerated plasma not directly associated with active jets \citep{Murgia2011, Brienza2017, Mahatma2018}, and restarted radio galaxies in which older remnant lobes are observed simultaneously with freshly-inflated radio lobes \citep{Schoenmakers2000, Jamrozy2007, Jetha2008a, Mahatma2019}. The morphological diversity of radio galaxies also encompasses hybrid morphology radio sources \citep{Gopal-Krishna2000, Kapinska2017, Mingo2019}, X-shaped radio galaxies \citep{Leahy1992, Yang2019a}, and more broadly, unclassified sources with unusual morphologies \citep{Gopal-Krishna2024a}. Amongst these unusual morphologies, radio galaxies can present plumed lobe characteristics, `disturbed' lobes (in contrast to `polar' lobes, which are in directly opposite directions), but also radio galaxies that do not exhibit clear jet-inflated lobes. 

There are many numerical studies \citep[see][for a review]{Bourne2023} investigating how AGN jets produce some of these morphologies  \citep[e.g.,][]{Saxton2010, Krause2012, Tchekhovskoy2016, Massaglia2016, Nolting2019a, ONeill2019, Mukherjee2020, Horton2020, Yates-Jones2023, Chen2023, Bhattacharjee2024, Elley2026, Dominguez-Fernandez2024}. In this study, we aim to combine three elements to simulate the morphology of radio galaxies in galaxy cluster environments. We describe them in the following:

\begin{enumerate}
\item Self-regulation: in cool-core cluster environments, central radio galaxies provide a heating component essential to regulate cooling flows. Simulating AGN jets that are able to self-regulate, that is where accretion onto the central SMBH directly powers the jets while maintaining the cluster in a cool-core state \citep[see e.g.][]{Dubois2010, Li2014, Prasad2015, Yang2016, Beckmann2019a, Wang2020, Bourne2021, Weinberger2026}, is an important component to reproduce realistic environments hosting central radio galaxies. Moreover, the potential for CR protons to heat the ICM makes them a compelling component to include in such simulations.

\item Varying jet luminosity: in contrast to single jet events often used in numerical works investigating AGN jet morphology, a self-regulating jet model naturally leads to an evolving jet luminosity, sometimes with interruptions in jet activity. Indeed, a varying jet power can have drastic effects on morphology \citep{Saxton2010, Whitehead2023a, Elley2026}. Most importantly, observations of remnant and restarted radio galaxies necessarily imply physical systems in which the AGN jets are not always active. Radio observations of filaments linked to radio galaxies \citep{Rudnick2022, Rubeis2025, Brienza2025} also point to complex interactions between jets and cluster environment \citep{Dubois2009, ZuHone2021, Bourne2017, Bellomi2024, Dominguez-Fernandez2024}.

\item Spectral modelling of CR electrons: any attempt to produce realistic radio galaxy morphology from numerical simulations should involve modelling the non-thermal electrons responsible for the emitted synchrotron radiation, and self-consistently evolved magnetic fields. The short cooling timescales of CR electrons \citep{Ensslin2011} necessitate the inclusion of spatially and temporally evolving radiative losses. Moreover, evolving resolved CR electron spectra rather than individual Lorentz factors entails the inclusion of processes such as Coulomb and bremsstrahlung losses, which can have an impact on the shape of obtained radio spectra \citep{Werhahn2026b}, and which are important to study fossil electron populations in galaxy clusters \citep{Pinzke2013, vanWeeren2019}.
\end{enumerate}

In this work, we present simulations combining these three elements: 1) jets providing a heating source with the inclusion of a CR proton component, stabilizing the cool-core cluster, 2) a time-varying jet luminosity determined by accretion onto the central SMBH, and 3) the spectral evolution of CR electrons responsible for the observed radio emission.

\citet{Ehlert2023} -- hereafter \citetalias{Ehlert2023} -- presented three-dimensional magneto-hydrodynamical (MHD) simulations of an idealised Perseus-like cool-core cluster in combination with the AGN jet model from \citet{Weinberger2023}. Their simulated cluster self-regulates and exhibits central entropies and cooling times on par with observations of cool-core clusters. Here, we make use of their cold accretion model, which mimics cold gas raining onto the central supermassive black hole \citep{Pizzolato2005, Sharma2012a, Gaspari2017, Tremblay2018, Olivares2019}. We extend their setup by implementing CR proton and electron populations, first presented in a simpler, single jet event in \citet{Jlassi2026a} -- hereafter, \citetalias{Jlassi2026a}. In this study, we focus our analysis on the spectral evolution of CR electrons and the resulting synthetic radio observations. The impact of CR protons on the thermodynamic state of the cluster will be tackled in future studies, for which heating estimators can be used \citep[e.g.][]{Martizzi2019, Meenakshi2026}.

We organise the paper as follows: in Sect.~\ref{sec:methods}, we describe the simulation setup and the employed CR models. In Sect.~\ref{sec:results:self-regulation-radio-mocks}, we present the mock radio observations of the self-regulating cool-core cluster. In Sect.~\ref{sec:results:global-electron-spectra}, we present the global CR electron and radio spectra throughout the simulation. Finally, in Sect.~\ref{sec:results:contributions-different-ages}, we examine the local spectra of various electron populations and how they contribute to global spectra.

\section{Methods}\label{sec:methods}
In this work, we use identical initial conditions, accretion and jet modelling to those of the {\fontfamily{lmtt}\selectfont Fiducial} model described in \citetalias{Ehlert2023}. We summarise all essential aspects here but refer the reader to their work for a more detailed description. The novel aspect of our simulations lies in the implementation of CR proton and electron populations and is described in Sects.~\ref{sec:CRps} and \ref{sec:CRes}.

Our three-dimensional MHD simulations employ the \textsc{Arepo} code \citep{Springel2010, Pakmor2016a, Weinberger2020}. It uses an unstructured, moving Voronoi mesh on which the equations of ideal MHD are discretised and solved using the second-order finite-volume Godunov scheme \citep{Pakmor2011, Pakmor2013} and an HLLD Riemann solver \citep{Miyoshi2010}.

\subsection{Initial conditions: a Perseus like cool-core cluster}\label{sec:ICs}
The simulated galaxy cluster is initialised in a cubic domain of side length 6 Mpc. The electron density follows the radial profile of \citet{Churazov2003}, rescaled to a cosmology with a dimensionless Hubble parameter of
$h=0.67$:
\begin{equation}
\begin{split}
n_{\rm{e}} & = 46 \times 10^{-3} \left[1 + \left( \frac{r}{60 ~\rmn{kpc}} \right)^2 \right]^{-1.8} \rmn{cm}^{-3} \\
& + 4.7 \times 10^{-3} \left[ 1 + \left( \frac{r}{210 ~\rmn{kpc}} \right)^2 \right]^{-0.87} \rmn{cm}^{-3},
\end{split}
\end{equation}
and the gas is placed in hydrostatic equilibrium within a composite gravitational potential. The potential consists of an NFW cluster halo with virial mass $M_{\rmn{200, NFW}} = 8 \times 10^{14} \, \rmn{M}_\sun$, radius $R_{\rmn{200,NFW}} = 2$ Mpc and concentration parameter 5 \citep{Reiprich2002}, along with a central galaxy represented by an isothermal sphere with mass $M_{\rmn{200, ISO}} = 2.4 \times 10^{11}  \, \rmn{M}_\sun$ and radius $R_{\rmn{200,ISO}} = 15$~kpc \citep{Mathews2006}. 

The initial conditions include turbulent magnetic fields using the procedure from \citet{Ruszkowski2007} and \citet{Ehlert2018} to yield an average magnetic-to-thermal pressure ratio of $X_{B,\rmn{ICM}} = P_{B} / P_\rmn{th} = 0.0125$, where $P_{B}$ and $P_\rmn{th}$ are the magnetic and thermal pressures, respectively. Perturbations are also added to the hydrostatic temperatures. In this way, both magnetic and temperature fluctuations follow a Kolmogorov power spectrum on scales smaller than $k_\rmn{inj} = (37.5 \rmn{\,kpc})^{-1}$, and white noise on larger scales. In addition, velocity fluctuations with a dispersion of $\sigma = 70$~km~s$^{-1}$ are imposed within the central $800$ kpc. Although this is lower than measured velocity dispersions \citep{HitomiCollaboration2018, Zhang2025}, these initial fluctuations are erased by subsequent jet activity. The fluctuations in the initial conditions render this cool-core cluster environment more realistic by adding asymmetry mainly through orbiting substructures and local sites where thermal instabilities can grow at different stages.

\subsection{Cooling and star formation}\label{sec:cooling-star-formation}
We model cooling and star formation consistent with the IllustrisTNG model \citep{Pillepich2018}, which built upon the model previously described in \citet{Vogelsberger2013}. Radiative cooling including primordial and metal-line cooling with constant metallicity $Z = 0.3 \, \rmn{Z}_{\rmn{\sun}}$ \citep{Werner2013} is modelled above temperatures of $10^4$~K. Gas denser than $n_{\rmn{H}} = 0.13 \, \rmn{cm}^{-3}$ either exists on an effective equation of state as star-forming gas, or as hotter, non star-forming gas. Star formation is modelled as a stochastic process calibrated to the Kennicutt-Schmidt relation according to the interstellar medium (ISM) model described in \citet{Springel2003}. In contrast to IllustrisTNG, we do not use wind particles for stellar feedback in our simulations.

\subsection{Accretion and AGN feedback}\label{sec:AGN-feedback}
In contrast to the fixed jet power used in our previous work \citepalias{Jlassi2026a}, we use the self-regulated AGN jet feedback presented in \citetalias{Ehlert2023}, where the jet power is determined by accretion onto the SMBH throughout the simulation duration of 2~Gyr.

In the centre of the cool-core cluster, we place a black hole of mass $M_\rmn{BH} = 4.5 \times 10^9 \, \rmn{M}_{\sun}$  based on black hole -- host galaxy scaling relations \citep{Kormendy2013}. Around this black hole, an inner sphere with radius $r = 0.65$~kpc is used to initialise the jet, while an outer spherical shell $0.65 < r < 2$~kpc is used to determine the accretion properties onto the black hole \citep[see][for details on the numerical implementation of the jet algorithm]{Weinberger2023}.

In the outer spherical shell ($0.65 < r < 2$~kpc), we make use of the cold accretion model implemented in \citetalias{Ehlert2023}, where the black hole accretion rate is defined as
\begin{equation}\label{eq:cold_accreted_mass}
	\dot{M}_{\rm{cold}} = \epsilon \frac{M_{\rm{cold}}}{t_{\rm{ff}}},
\end{equation}
where $M_{\rm{cold}}$ includes gas with a temperature below $2 \times  10^4$ K and star forming gas, $t_{\rm{ff}} = \sqrt{2r/g}$ is the free-fall time with $g = \mathrm{d}\Phi / \mathrm{d} r$ being the local acceleration due to the composite gravitational potential described in Sect.~\ref{sec:ICs}. The parameter $\epsilon$ includes effects unresolved in our simulations that can affect the accretion of cold gas, namely angular momentum delaying accretion of the gas onto the SMBH and unresolved small-scale feedback heating some of the cold gas before it is accreted.

The jet power is determined by the accretion rate according to
\begin{equation}\label{eq:jet_power}
	L_{\rm{jet}} = \eta \dot{M}_{\rm{cold}} c^2 = \eta \epsilon \frac{M_{\rm{cold}}}{t_{\rm{ff}}}  c^2,
\end{equation}
where $c$ is the speed of light and $\eta$ is an efficiency parameter. Following the {\fontfamily{lmtt}\selectfont Fiducial} simulation of \citetalias{Ehlert2023}, we use efficiencies $\epsilon = 0.1$ and $\eta = 0.01$. More details on the influence of accretion model (Bondi vs. cold accretion) on the self-regulated nature of the cool-core cluster can be found in \citetalias{Ehlert2023}.

We initialise the jets within the central spherical region $r < 0.65$~kpc with a fiducial jet density of  $\rho_{\rmn{jet}} = 10^{-28} \,$~g~cm$^{-3}$ corresponding to an ICM-to-jet density contrast of $\rho_{\rmn{ICM}} / \rho_{\rmn{jet}} \sim 3 \times 10^3-10^4$. The jets are launched from this region in strictly bipolar directions with zero opening angle. Additionally, toroidal magnetic fields are injected by setting the magnetic-to-thermal pressure ratio $X_{B, \rm{jet}} = 0.1$. Jet material is traced using a passive scalar initialised to $X_{\rmn{jet}} = 1$ in cells located in the jet region, and subsequently advected with the flow.

The cell mass resolution is determined using the distance from the centre following $m_\rmn{target} = m_\rmn{target,0} \exp\left({\frac{r}{100 \,\rmn{kpc}}}\right)$, where  $m_\rmn{target,0} = 1.5 \times 10^6 \, \rmn{M}_\sun$. At the cluster outskirts, the cell volume is limited to $V^{1/3}_\rmn{cell} \approx 370 \,$~kpc for numerical reasons. Jet cells with $X_{\rmn{jet}} > 10^{-3}$ are refined to a target volume $(V_\rmn{jet,\, target}/X_{\rmn{jet}})^{1/3}$ with $V_\rmn{jet,\, target} = 0.65$~kpc, but at most a volume where a cell does not exceed the target mass $m_\rmn{target}$. This ensures that the jet propagation is well-resolved, while smoothly derefining cells with aged jet material to the target mass criterion of the ICM. Cells are refined (derefined) if the target mass or volume exceeds (is below) a factor of 2 of the target volume. On top of these criteria, we refine cells to ensure that adjacent cells have volumes differing at most by a factor of four.

\subsection{Cosmic ray protons}\label{sec:CRps}
In jets, CR acceleration can occur through shocks at re-collimation and termination sites \citep{Nishikawa2020, Cerutti2023}, backflows \citep{Matthews2019} or shearing flows where reconnection can occur \citep{WangJ2021, Sironi2021}. Our AGN feedback model is aimed at studying the long-term impact of AGN jets on cool-core galaxy clusters ($>$~Gyr). Consequently, due to resolution limitations, the sites of acceleration are unresolved in our simulations. We therefore employ a sub-grid prescription for acceleration of CRs, which we describe in detail in \citetalias{Jlassi2026a}. We summarise the main aspects here.

CR protons are evolved as a relativistic fluid with adiabatic index $\gamma = 4/3$ and dynamically coupled to the total pressure in the MHD equations following \citet{Pfrommer2017}. The energy converted into CR protons is determined based on the jet energy according to
\begin{equation}\label{eq:fraction_jet_crp}
E_{\rm{crp}} = \xi_{\rm{crp}} E_{\rm{jet}},
\end{equation}
where $E_{\rm{crp}}$ and $E_{\rm{jet}}$ are the CR proton and jet energies, and $\xi_{\rm{crp}} = 0.1$ is the efficiency of jet-to-CR proton energy conversion. Converting thermal energy into CR protons during jet launch would alter jet dynamics. To avoid this, thermal energy is instead gradually converted into CR proton energy inside the jet, after the jet's kinetic energy has thermalised (this occurs faster than the dynamics of the system). We keep track of the energy budget to be converted progressively into CR protons using a passive quantity $\mathcal{E}_{\rmn{crp}}$, advected in \textsc{Arepo}. The evolution of this energy budget is dictated by an exponential decay
\begin{equation}
\frac{\rmn{d}\mathcal{E}_{\rmn{crp}}}{\rmn{d}t} = -\frac{\mathcal{E}_{\rmn{crp}}}{\tau_{\rmn{inj}}}
\end{equation}
over a characteristic injection timescale $\tau_{\rmn{inj}} = 10$~Myr, chosen to be comparable to the duration the relativistic material remains in the jet spine. In our simulations where the jet power is dictated by accretion (and therefore stochastic, see Sect.~\ref{sec:AGN-feedback}), using $\mathcal{E}_{\rmn{crp}}$ to track the energy budget to be converted into CRs allows us convert the correct fraction of jet energy into CR protons. Along a Lagrangian element of the jet, our algorithm leads to an accelerated CR energy that exponentially decreases with time, as described in detail in \citetalias{Jlassi2026a}. This injected energy also serves as the source function for CR electron acceleration.

In addition to advection with the background plasma, CRs are transported through diffusion and streaming. When CRs propagate faster than the Alfv\'en velocity, they excite Alfv\'en waves: this process is the streaming instability \citep{Kulsrud1969,Shalaby2023,Lemmerz2025}. As CRs scatter on their self-generated Alfv\'en waves, they start to isotropize in the wave frame. Damping processes dissipate wave energy and hence, effectively transfer energy from CRs to the background thermal plasma following a rate $| \bm{\varv}_{\rmn{A}} \bm\cdot \bm\nabla P_{\rmn{crp}}|$ \citep{WienerEtAl2013} where $\bm{\varv}_{\rmn{A}} = \bm{B} /\sqrt{4 \pi \rho}$ is the Alfv\'en velocity, $\rho$ is the mass density and $\bm\nabla P_{\rmn{crp}}$ is the CR pressure gradient.

We model anisotropic diffusion along the local magnetic field with a constant anisotropic diffusion coefficient along the magnetic field of $\kappa_{\rmn{diff}} = 10^{29} \, \rmn{cm}^2 \, \rmn{s}^{-1}$. To order of magnitude, this corresponds to an Alfv\'en speed of $100\,\rmn{km \, s}^{-1}$ and a CR gradient length of 3~kpc. The combination of this anisotropic diffusion with Alfv\'en losses emulates the effect of streaming. Alfv\'en wave losses are only active in cells with $X_{\rmn{jet}} \leq 10^{-3}$ to prevent artificially large numerical cooling rates at steep gradients in interfaces between lobes and the surrounding ICM \citep[we point the reader to][for further details]{Ehlert2018}. Furthermore, we include Coulomb and hadronic losses \citep{Pfrommer2017} of CR protons, which are subdominant in comparison to Alfv\'enic losses in galaxy clusters \citep{Guo2008, Ruszkowski2017}.

Appendix~\ref{fig:radprof_thermodynamics} shows the radial evolution of the CR proton to thermal pressure ratio $X_{\rmn{crp}}$$X_{\rmn{crp}}$, and also confirms that our cool-core cluster self-regulates and exhibits a stable thermodynamic state across 2~Gyr with the addition of this CR proton population, in agreement with \citetalias{Ehlert2023}. In a future study, we will investigate the impact of CR protons on a self-regulated cool-core such as the one presented here, in comparison to one without CRs \citepalias{Ehlert2023} using a heating estimator \citep[e.g.][]{Meenakshi2026}. This will be crucial to quantify the contribution from different heating components (e.g. CRs, shocks, turbulent dissipation and mixing, adiabatic compression).

\subsection{Cosmic ray electrons}\label{sec:CRes}
The modelling of CR electron spectra is performed in the Lagrangian frame using tracer particles, which track fluid properties on the MHD timestep \citep[these are the velocity fluid tracers described in][]{Genel2013}. Along these Lagrangian trajectories, the post-processing code \textsc{Crest} \citep{Winner2019} solves the Fokker-Planck equation to evolve the one-dimensional CR electron distribution function $f(p) = 4 \pi p^2 f^{\rmn{3D}}(p)$, where $f^{\rmn{3D}}(p)$ is the three-dimensional distribution function, and $p = |\bm{p}|  = |\bm{P}| / (m_\rmn{e} c)$ is the normalised momentum, where $m_{\rmn{e}}$ is the electron mass. Details of the numerical methods and physics implemented in this code are described in \citet{Winner2019}.

We evolve the distribution function $f(p)$ accounting for sub-grid acceleration in AGN jets \citepalias{Jlassi2026a}, Coulomb collisions, synchrotron, inverse Compton, and bremsstrahlung losses, and adiabatic changes. We developed a specific procedure to treat mixing effects for low-density AGN jets in dense cool-core galaxy cluster environments \citepalias{Jlassi2026a}, which we summarise in Sect.~\ref{sec:adiabatic-effects}. \textsc{Crest} is additionally equipped to perform sub-grid acceleration at supernovae \citep{Werhahn2026a}, resolved Fermi-I acceleration \citep{Whittingham2026a, Whittingham2026b} and re-acceleration, and Fermi-II momentum diffusion, which are not modelled in this work. 

For each tracer particle, which represents a population of CR electrons, we obtain a temporally and spatially evolving electron spectrum. The CR electron spectrum can be integrated to yield the CR electron energy density $\varepsilon_{\rmn{cre}} = \int_0^\infty T_{\rmn{e}}(p) f(p) \rmn{d}p$ where $T_{\rmn{e}}(p) = (\sqrt{1 + p^2} - 1)m_{\rmn{e}} c^2$ is the CR electron kinetic energy. \textsc{Crest} therefore samples the underlying CR electron density field. A Voronoi mesh is constructed in post-processing based on the location of the tracer particles (i.e. a different tessellation from the one constructed in \textsc{Arepo}), allowing each tracer to be associated with a volume $V_{\rmn{cell,tr}}$, and hence an energy according to $E_{\rmn{cre}} = \varepsilon_{\rmn{cre}} V_{\rmn{cell,tr}}$ (see Fig.~\ref{fig:energies}).

We integrate $f(p)$ in the range from $10^{-2} < p < 10^{8}$ with 20 logarithmically-spaced bins per decade. We consider the ICM to be a fully ionised primordial gas containing primarily hydrogen with a mass fraction of $X_{\rm{H}} = 0.76$, and helium. This leads to a mean molecular weight of $\mu = 0.588$ and an electron-to-hydrogen abundance of $x_{\rmn{e}} = 1.157$. We use a redshift of $z = 0$, corresponding to a cosmic microwave background (CMB) photon energy density of  $\varepsilon_{\rmn{cmb}} \approx 4.172 \times 10^{-13}\, \rm{erg} \, \rm{cm}^{-3}$, or equivalently, a magnetic field strength $B_{\rmn{cmb}} = 3.2 \, \mu \rmn{G}$. Although we use initial conditions based on the Perseus galaxy cluster (see Sect.~\ref{sec:ICs}), this work does not aim to make specific predictions for this particular object. Rather, this object serves as an archetype cool-core galaxy cluster in which AGN feedback is at play. Using a redshift of $z = 0.0183$ would lead to a marginal increase in CMB equivalent magnetic field strength at this redshift to  $B_{\rmn{cmb}} = 3.2 \, (1+z)^2\, \mu \rmn{G} \approx 3.3\, \mu \rmn{G}$. We therefore only use this redshift to convert observational beam sizes to physical distances in Fig.~\ref{fig:radio_maps}.

Tracer particles are spawned on-the-fly in the jet injection region. Rather than creating a tracer particle in each cell in the jet region, which could lead to an unpredictably large number of tracer particles due to the stochasticity of jet activity in our simulations, we create tracer particle at a given frequency $1 / T_{\rmn{tracer \, creation}}$, where $T_{\rmn{tracer \, creation}}$ is the time period between tracer creation events. We choose the latter according to the final number of tracers we aim to spawn during the simulation. 
Additionally, we add a background distribution of tracer particles\footnote{These background tracers do not contribute to the total spectrum, as they only get accelerated if they are entrained in the jet. They serve to compute the Voronoi mesh in post-processing by filling regions devoid of tracers.} in the initial conditions filling a cubic box of side 340~kpc. In the simulations presented here, we obtain a total of $6 \times 10^5$ tracers at the end of the simulation. This corresponds to one tracer particle for every two cells with $X_{\rmn{jet}} > 10^{-6}$.

\subsubsection{Acceleration of CR electrons}\label{sec:cre-acceleration}

Following the procedure initially described in \citetalias{Jlassi2026a}, we convert a fraction of the energy density accelerated into CR protons (Sect.~\ref{sec:CRps}) into CR electrons according to
\begin{equation}\label{eq:fraction_crp_cre}
\varepsilon_{\rm{cre}} = \xi_{\rm{cre}} \varepsilon_{\rm{crp}},
\end{equation}
where $\xi_{\rm{cre}} = 0.01$ is the energy conversion efficiency from CR protons to electrons. Combining proton (see Sect.~\ref{sec:CRps}) and electron acceleration efficiencies results in 0.1~\% of the jet energy being converted into CR electrons. We only accelerate CR electrons when they are inside the jets or lobes, defined here as $X_{\rmn{jet}} > 10^{-3}$ and when the gas speed exceeds $\varv_{\rmn{min}} = 3000 \, \rmn{km}\,\rmn{s}^{-1}$. We model the accelerated CR electron population as a power-law in momentum with a source function following
\begin{equation}
Q_{\rmn{inj}}(p) = r_{\rm{jet}} \frac{C}{\Delta t} p^{-\alpha_{\rmn{inj}}} \Theta (p - p_{\rmn{min}}),
\label{eq:injected_spectrum}
\end{equation}
where $\alpha_{\rmn{inj}} = 2.2$ is the electron injection spectral index. Given a synchrotron flux density $S\propto \nu^{-\alpha_\nu}$ where $\nu$ is the frequency, this electron injection index corresponds to a radio spectral index of $\alpha_{\nu} = (\alpha_{\rm{inj}} - 1) /2 = 0.6$, as observed in radio lobes \citep[e.g.][]{Murgia2011, Kolokythas2015, Shulevski2017, Mahatma2023}. The variable $r_{\rm{jet}}$ is required to account for mixing effects (see Sect.~\ref{sec:adiabatic-effects}), $\Delta t$ is the MHD timestep, and $\Theta(x)$ is the Heaviside step function. The normalisation of the spectrum $C$ and the minimum injection momentum $p_{\rmn{min}}$ are determined based on the accelerated CR electron energy density and enforcing continuity between the thermal Maxwellian and the non-thermal CR electron spectrum. We limit $3 \, p_{\rmn{th}}\leq p_{\rmn{min}} \leq 10$ \citepalias{Jlassi2026a}, where $p_{\rmn{th}}$ is the momentum at which the thermal Maxwellian peaks.

\subsubsection{Adiabatic effects and mixing}\label{sec:adiabatic-effects}

Low-density, high-temperature jet plasma injected into the dense, cold central parts of cool-core clusters leads to a convectively unstable configuration. As a consequence, jet-inflated bubbles rise buoyantly in the cluster atmosphere, thereby expanding and hence adiabatically cooling. Along Lagrangian trajectories of tracer particles created in low-density jets, mixing of low-density jet material with the dense ICM causes the density $\rho$ recorded by tracers to increase in time. This increase in density also causes adiabatic compression of CR electrons. Simultaneously, however, as jet material mixes with the ICM, CR electrons initially injected in jet plasma adiabatically expand into a larger volume devoid of electrons.

We treat adiabatic compression and expansion of the spectrum using the background density $\rho$ recorded by tracer particles. In addition, we account for the second effect, namely advection and mixing of low-density jet material (filled with CR electrons) with high-density ICM material (devoid of CR electrons), by using the jet tracer $X_{\rm{jet}} = m_{\rmn{jet}} /(m_{\rmn{ICM}} + m_{\rmn{jet}})$, where $m_{\rmn{ICM}}$ and $m_{\rmn{jet}}$ are the masses of ICM and jet materials, respectively. We make use of $X_{\rmn{jet}}$, which is evolved using the continuity equation and therefore captures mass flow across cell interfaces accurately, to quantify how much CR electrons have mixed with the ICM. Specifically, we use this quantity to track changes in the number density of electrons. Indeed, CR electrons initially accelerated in the jet where $X_{\rmn{jet}} = 1$ experience dilution as jet material mixes with the ICM, causing $X_{\rmn{jet}}$ to decrease. In practice, this is taken into account in the CR electron spectrum by rescaling the number density defined as $n_{\rmn{cre}} = \int_0^\infty f(p) \rmn{d}p$, according to
\begin{equation}\label{eq:dilution}
n_{\rmn{cre, f}} = r_{\rm{jet}}  n_{\rmn{cre, i}},
\mbox{ where }
r_{\rm{jet}} = X_{\rm{jet, f}}/X_{\rm{jet, i}},
\end{equation} 
where the subscripts $\rm{i}$ and $\rm{f}$ refer to a change from an initial state to a final state. This is simply a change in the normalisation of the spectrum, $f_{\rmn{f}} = r_{\rm{jet}} f_{\rmn{i}}$, following the definition of the CR electron number density. In \citetalias{Jlassi2026a}, we investigated adiabatic and mixing effects in great detail. Overall, we found that CRs accelerated by low-density jets in cool-core environments experience both adiabatic compression and dilution, leaving the overall normalisation unchanged but shifting spectral distributions to higher momenta.

\subsection{Synchrotron emission}\label{sec:synchrotron}

To obtain synthetic synchrotron data, we employ the \textsc{Crayon+} code \citep{Werhahn2021c}, which post-processes individual tracer particles from \textsc{Crest} to calculate the instantaneous non-thermal emission. We point the reader to Appendix~A.1 in \citealt{Werhahn2021c} for more details.
In the following, we assume that the synchrotron emission is optically thin to self-absorption in the ICM. We calculate the synchrotron emissivity for each tracer particle as (in units of erg~s$^{-1}$~Hz$^{-1}$~cm$^{-3}$)
\begin{equation}
j_\nu = E \frac{\rmn{d}N_\gamma}{\rmn{d}\nu \rmn{d}V \rmn{d}t} = \frac{\sqrt{3} e^3 B_\perp}{m_\rmn{e} c^2} \int^\infty_0 f(p) F(\nu / \nu_\rmn{c}) \, \rmn{d}p,
\end{equation}
where $E$ is the energy of $N_\gamma$ photons, $\nu$ is the frequency, $V$ is the unit volume, $t$ is the unit time, $e$ is the elementary charge, and $B_\perp$ is the magnetic field perpendicular to the line of sight. The dimensionless synchrotron kernel $F(x)$ in the integral is defined as 
\begin{equation}
F(x) = x \int^\infty_0 K_{5/3}(\xi) \rmn{d}\xi,
\end{equation}
where $x = \nu / \nu_\rmn{c}$ and $K_{5/3}$ is the modified Bessel function of order 5/3. An analytical approximation from \citet{Aharonian2010} is used to calculate $F(x)$ more efficiently .

Integrating $j_\nu$ along the line of sight $L$ allows us to obtain the radio synchrotron intensity $I_\nu$ (in units of erg~s$^{-1}$~Hz$^{-1}$~cm$^{-2}$~sterad$^{-1}$)
\begin{equation}
I_\nu = \frac{1}{4 \pi} \int^\infty_0 j_\nu \,\rmn{d}L,
\end{equation}
where we use $L = 340$~kpc for radio intensity maps throughout this work, unless otherwise specified.
The synchrotron spectral index is then calculated according to
\begin{equation}\label{eq:spectral_index}
\alpha_{\nu_1}^{\nu_2} = - \frac{\log_{10}(I_{\nu_2}/I_{\nu_1})}{\log_{10}(\nu_2 / \nu_1)},
\end{equation}
where $\nu_1 < \nu_2$ such that $\alpha_{\nu_1}^{\nu_2} > 0$.
The critical frequency $\nu_{\rmn{c}}$ is
\begin{equation}
\nu_\rmn{c} = \frac{3 e B_\perp \gamma^2}{4 \pi m_\rmn{e} c},
\end{equation}
where $\gamma = \sqrt{1+ p^2}$ is the electron Lorentz factor, and $\gamma \approx p$ for $p \gg 1$. Given the typical synchrotron emission frequency $\nu_\rmn{sync} \approx 2 \nu_c$ \citep{Pfrommer2022}, we can therefore define the typical electron momentum emitting synchrotron radiation at 150~MHz as
\begin{equation}\label{eq:p-contributing}
p|_{150 \, \rmn{MHz}} \simeq 1.3 \times 10^3 \, \times \left( \frac{B_{\rmn{\perp}}}{10 \, \mu \rmn{G}} \right)^{-1/2},
\end{equation}
which is essential to understand how CR electron spectra contribute to the observed radio emission, as we investigate in Sect.~\ref{sec:results:contributions-different-ages}.

\section{Mock radio observations of self-regulated AGN jets in a cool-core cluster}\label{sec:results:self-regulation-radio-mocks}

\subsection{Jet deflection}\label{sec:results:jet-deflection}
\begin{figure*}[ht]
	\centering
	\includegraphics[width=2\columnwidth]{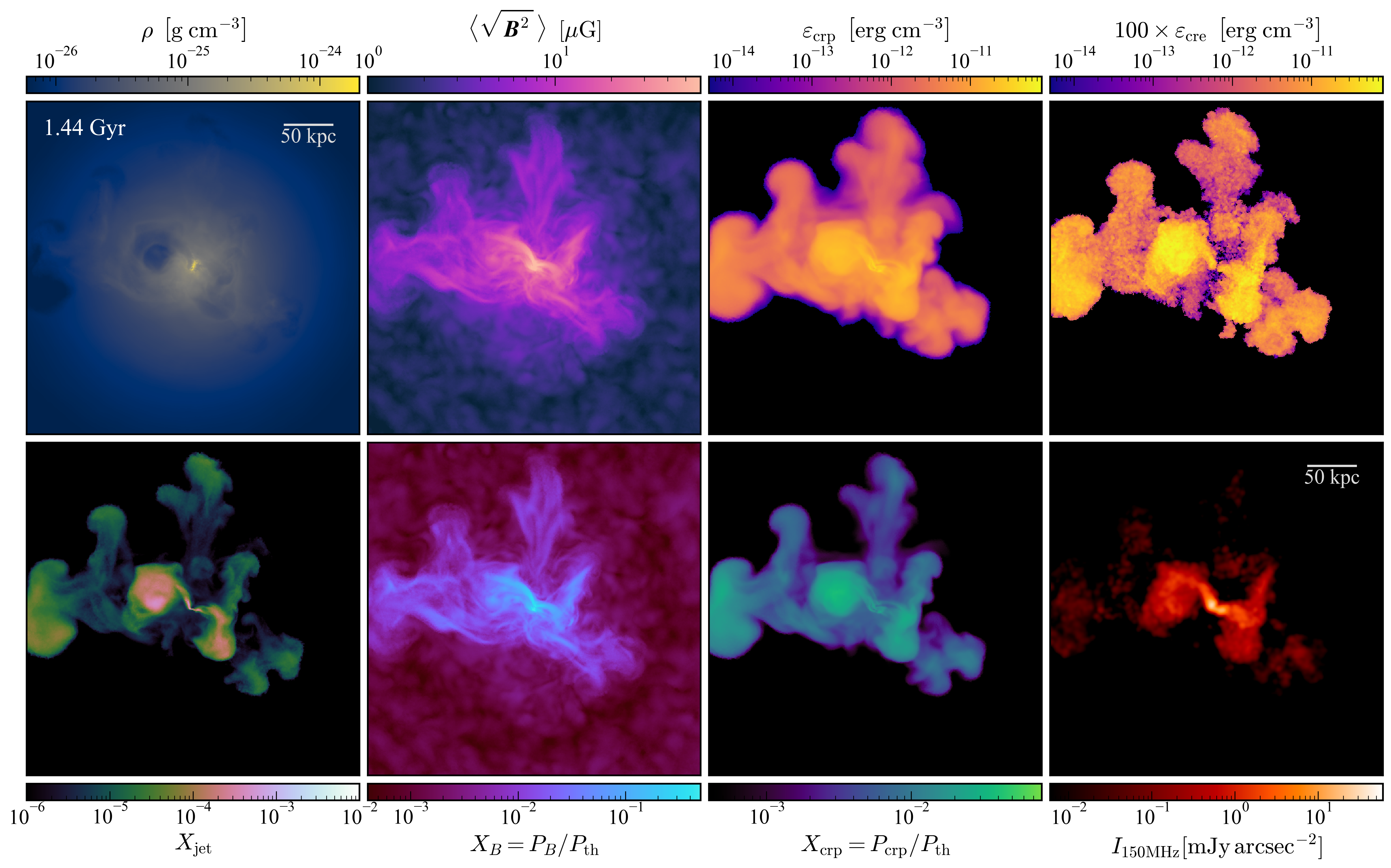}
	\caption{Projections of self-regulated AGN jets in a Perseus-like galaxy cluster at $t = 1.44$~Gyr. \textit{Top:} From left to right, volume-weighted gas mass density, magnetic field strength, CR proton energy density, and the CR electron energy density multiplied by $1/\xi_{\rm{cre}} = 100$, where $\xi_{\rm{cre}}$ is our chosen efficiency of CR proton to CR electron energy conversion (Sect.~\ref{sec:cre-acceleration}).
	\textit{Bottom:} From left to right, we show the volume-weighted jet tracer $X_{\rmn{jet}}$, the magnetic-to-thermal pressure ratio $X_{B}$, and the CR proton to thermal pressure ratio $X_{\rmn{crp}}$, all calculated using volume-weighted pressures. The last panel shows the radio intensity at 150~MHz obtained from CR electron spectra calculated with the Fokker-Planck solver \textsc{Crest}, smoothed with a 2D Gaussian beam with full-width at half-maximum $\rm{FWHM} = 4$~kpc.
	The low-density jet-inflated lobes of different shapes and sizes are filled with CRs. The radio intensity at 150~MHz primarily highlights the ongoing jet outburst, but previous outbursts are visible in the CR distributions.}
	\label{fig:gas_projections}
\end{figure*}

Figure~\ref{fig:gas_projections} shows projections of our simulations during a jet-active phase, at $t = 1.44$~Gyr. All projections have a depth $\approx 120$~kpc, with the exception of the radio intensity map (bottom right panel) which is obtained by projecting along a line of sight of $340$~kpc centred on the SMBH. The projection direction for all the panels is parallel to the jet launching direction.

Despite the jets being injected in a purely bi-directional manner with no opening angle and no jet precession, the diversity of lobe orientations echoes the strong jet deflection occurring on small scales, and is reflected in the $X_{\rmn{jet}}$ map on scales of 20~kpc close to the SMBH. The reason for this lack of collimation on large scales is due to a combination of two factors: one being intrinsic to the jet, and the other caused by the ambient medium. First, we simulate light jets with a density contrast of $\sim 3 \times 10^3-10^4$ with the external medium, which have insufficient momentum flux to push through the external medium. Second, we set up a turbulent ambient medium (see Sect.~\ref{sec:ICs}) where radiative cooling of the thermal gas leads to cold gas structures, which deflect the jet on kpc-scales. Indeed, light jets contribute to isotropic spreading of turbulence in comparison to denser jets (\citetalias{Ehlert2023}, \citealt{Meenakshi2026}). This turbulence is imprinted in the magnetic field strength map. Further away from the central AGN, the turbulent initial conditions are visible as local variations in the magnetic field strength. On the other hand, in the central parts, specifically in the jet and lobe regions, the magnetic field strength ranges from a few $\mu$G to $60 \, \mu$G, and exhibits filamentary structures along the jets but also in the wake of older, buoyantly rising lobes. These structures are also clear in the magnetic-to-thermal pressure ratio $X_{B}$ map, and appear more highly magnetised than the background gas\footnote{The influence of magnetic fields on cold gas was examined in detail in \citetalias{Ehlert2023}, who found that light jets lead to filamentary structures which are observable through H$\alpha$ in central dominant (cD) galaxies \citep[e.g.][]{Gendron-Marsolais2018}, as opposed to dense jets which form disk-like structures.}. This magnetic pressure support, reaching close to 50\% of the thermal pressure in the bubble wakes, provides an explanation for the filamentary morphology visible on larger scales. In Sect.~\ref{sec:results:contributions-different-ages}, we will discuss the specific impact of magnetic field strength on the resulting radio emission.

A variety of lobes from different jet outbursts can be seen in the jet tracer $X_{\rmn{jet}}$ projection. This passive quantity is initialised as $X_{\rmn{jet}} =1$ in the jet region, and decreases due to advection and turbulent mixing with ambient ICM gas, while adiabatic processes leave this quantity invariant. Because of this, smaller values of the jet tracer $X_{\rmn{jet}}$ (pink colours) indicate younger, more recently injected jet material\footnote{It is worth noting that the large projection depth needed to highlight the different lobes inflated in many directions, also lowers this volume-weighted quantity.}. Our self-regulated AGN jets produce diverse bubble shapes and sizes, located at different radii in almost all directions around the SMBH. In the mass density projection, these lobes are observable as under-dense regions, due to the low-density jets which inflate them. While the most recently inflated bubbles are clearly visible in the central 100~kpc, most of the older lobes at larger radii are less noticeable in the density projection. This is due to the large depth of projection used, but also due to the drop in ICM density towards larger radii, against which under-dense bubbles are less discernible. This diverse population of jet-inflated lobes lead to cavities in maps of synthetic X-ray emission, and will be examined in conjunction with our synthetic radio maps in future work.

\subsection{Self-regulation with CR protons and electrons}\label{sec:result:self-regulation-with-cr-protons-electrons}
This current work focuses on the new CR proton and electron populations implemented in this self-regulated cool-core environment, which are presented in the respective energy density maps in Fig.~\ref{fig:gas_projections}. Both CR protons and electrons are accelerated in the jet and hence reflect the distribution of jet material. However, while the jet material is simply advected, CR protons additionally experience diffusion, leading to smoother edges around the lobes. Turning our attention to the CR electron spatial distribution confirms that we obtain electron energy densities at the correct proportion to protons (i.e. smaller by a factor 100), although there are evident differences between the two: electrons retain slightly higher energy densities in the jet and lobes, and have more defined lobe edges as well as patchier distributions in the lobes. This reflects a difference in the modelled transport processes for both populations, as well as differing numerical implementations. Protons are evolved as a relativistic fluid in the finite-volume solver with advection, anisotropic diffusion in a turbulent magnetic field and dynamical pressure coupling, while electrons are evolved in post-processing along Lagrangian trajectories without accounting for diffusive transport. 

Moreover, protons and electrons experience different physical loss processes: their energies are therefore not expected to evolve in exactly the same manner, as shown by the total energy diagnostics presented in Appendix \ref{sec:appendix:self-regulation-total-energetics}. Yet the close spatial correspondence between the CR proton and electron distributions and that of the jet material indicates that diffusion of protons down the CR pressure gradient that is projected on the magnetic field lines is not the dominant transport mechanism. Instead, turbulent diffusion plays a large role in redistributing material in the cluster core, with a turbulent diffusion coefficient that can be approximated as $\kappa_{\rmn{turb}} \approx \sigma_{\rmn{los}} L_{\rmn{turb}}/3 \approx 10^{30} \, \rmn{cm}^2 \, \rmn{s}^{-1}$, where $\sigma_{\rmn{los}} \approx 180 \, \rmn{km} \, \rmn{s}^{-1}$ is the estimated line of sight velocity dispersion \citepalias[see Fig.~14 of][]{Ehlert2023} and the coherence scale is on the order of the diameter of jet-inflated bubbles, $L_{\rmn{turb}} \approx 60$~kpc \citep{Bellomi2025}. This effective turbulent diffusion coefficient is thus larger than the intrinsic CR diffusion coefficient in three-dimensional (nearly isotropic) turbulence of $\kappa_{\rmn{diff}}/3 \approx 3\times 10^{28} \, \rmn{cm}^2 \, \rmn{s}^{-1}$ (see Sect.~\ref{sec:CRps}).

The isotropic bubble distribution due to jet deflection leads to CRs filling the central 100~kpc of the cool-core cluster. The CR proton-to-thermal pressure map mirrors this, and additionally illustrates the effect of our CR proton algorithm in converting jet energy into CR protons with a varying jet power. We reach values of $X_{\rmn{crp}} \leq 0.05$ in the central 10~kpc and obtain lower values in the central $\sim$100~kpc after 1.5~Gyr (see Fig.~\ref{fig:radprof_thermodynamics}), which is in line with upper limits on  $X_{\rmn{crp}}$ derived from the non-detection of gamma-rays \citep{Jacob2017a, Jacob2017}. We defer further analysis on this topic to a future study.

Finally, the radio intensity map at 150~MHz highlights the central lobes resulting from the ongoing jet outburst, in which the jets appear bent on scales of 20~kpc, reminiscent of FRI jets which do not remain collimated on large scales \citep{DeGasperin2012, Wezgowiec2024, Cotton2025}. Beyond the central pair of lobes reaching ${\sim} 70$~kpc from the central SMBH, other lobes recognizable in the CR distributions have radio intensities lower by two orders of magnitude or more. However, the magnetic field strengths and CR electron energy densities in these lobes at larger radii have not decreased significantly. This suggests that older electrons are filling these lobes, which have cooled by the time they reach those radii. We discuss this in detail in Sect.~\ref{sec:results:age-distributions-radio-spectra}. To understand how these radio intensities fit within current observations of radio galaxies, we now present mock radio observations of our simulations at multiple epochs.

\begin{figure*}[ht]
	\centering
	\includegraphics[width=2.\columnwidth]{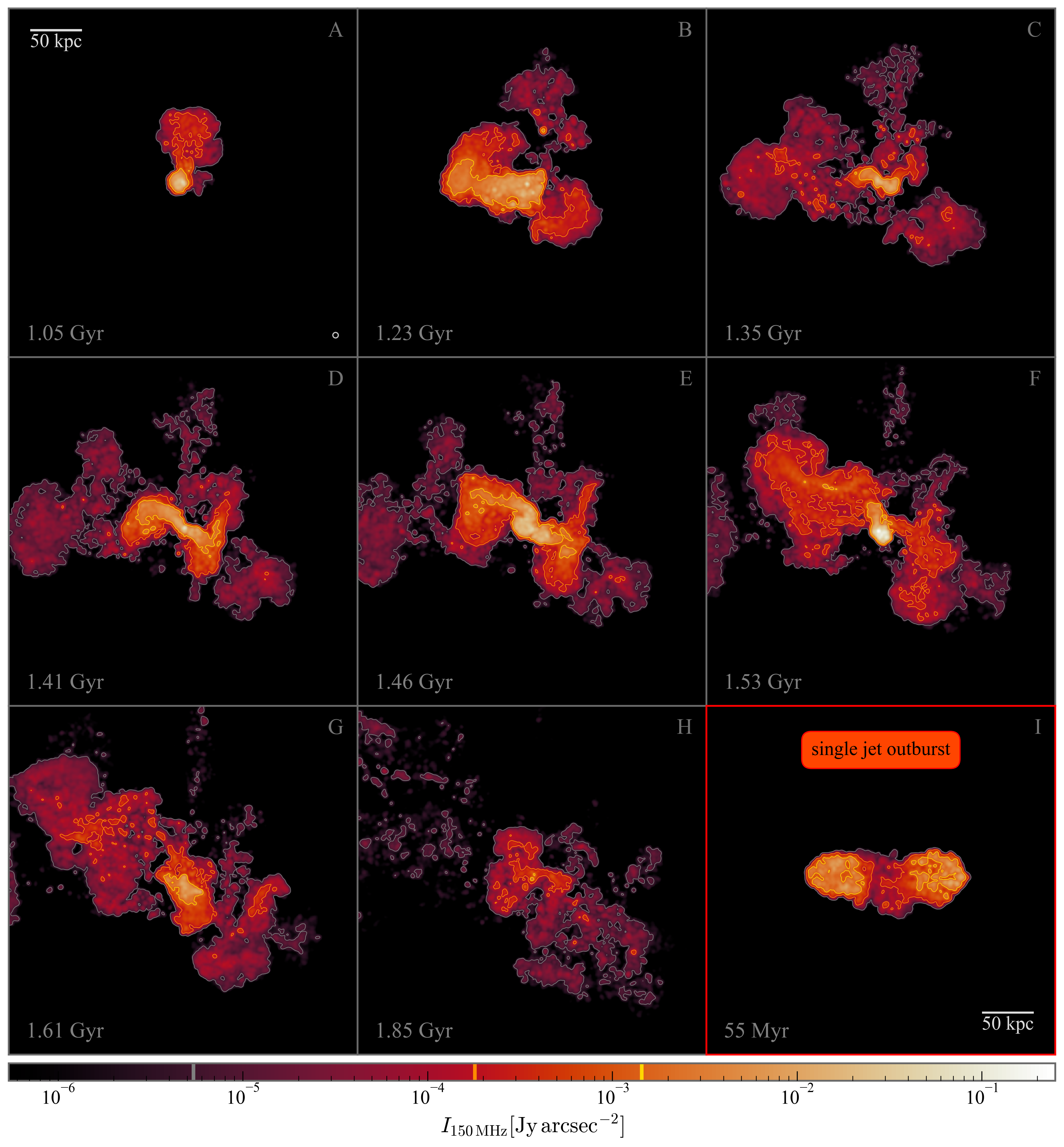}
	\caption{Synchrotron intensity maps at $150$~MHz of self-regulated AGN jets at different epochs, in comparison with the single jet outburst in the bottom right corner presented in \citetalias{Jlassi2026a}. We smooth the image using a symmetric 2D beam with FWHM $= 7\arcsec \approx 2.7 \, \rmn{kpc}$ (depicted in panel A at the bottom right) to mock observationally motivated angular resolutions, and use the conversion $1 \, \rmn{Jy}\,\rmn{beam}^{-1} = 0.018 \,\rmn{Jy}\,\rmn{arcsec}^{-2}$. We draw contours at $\sigma_{\rm{LoTSS}} \times [3, 100, 800]$, where $\sigma_{\rm{LoTSS}} = 100 \, \mu \rm{Jy / beam}$ is the sensitivity achieved in LoTSS. These contour levels are also marked in the colour bar as vertical lines. We obtain mock radio maps of FRI-like jets, and lobes with complex, asymmetrical structures. This in contrast to the simulation of a single jet outburst, which solely produces symmetrical, polar lobes.}
	\label{fig:radio_maps}
\end{figure*}

\begin{figure}[ht]
	\centering
	\includegraphics[width=1\columnwidth]{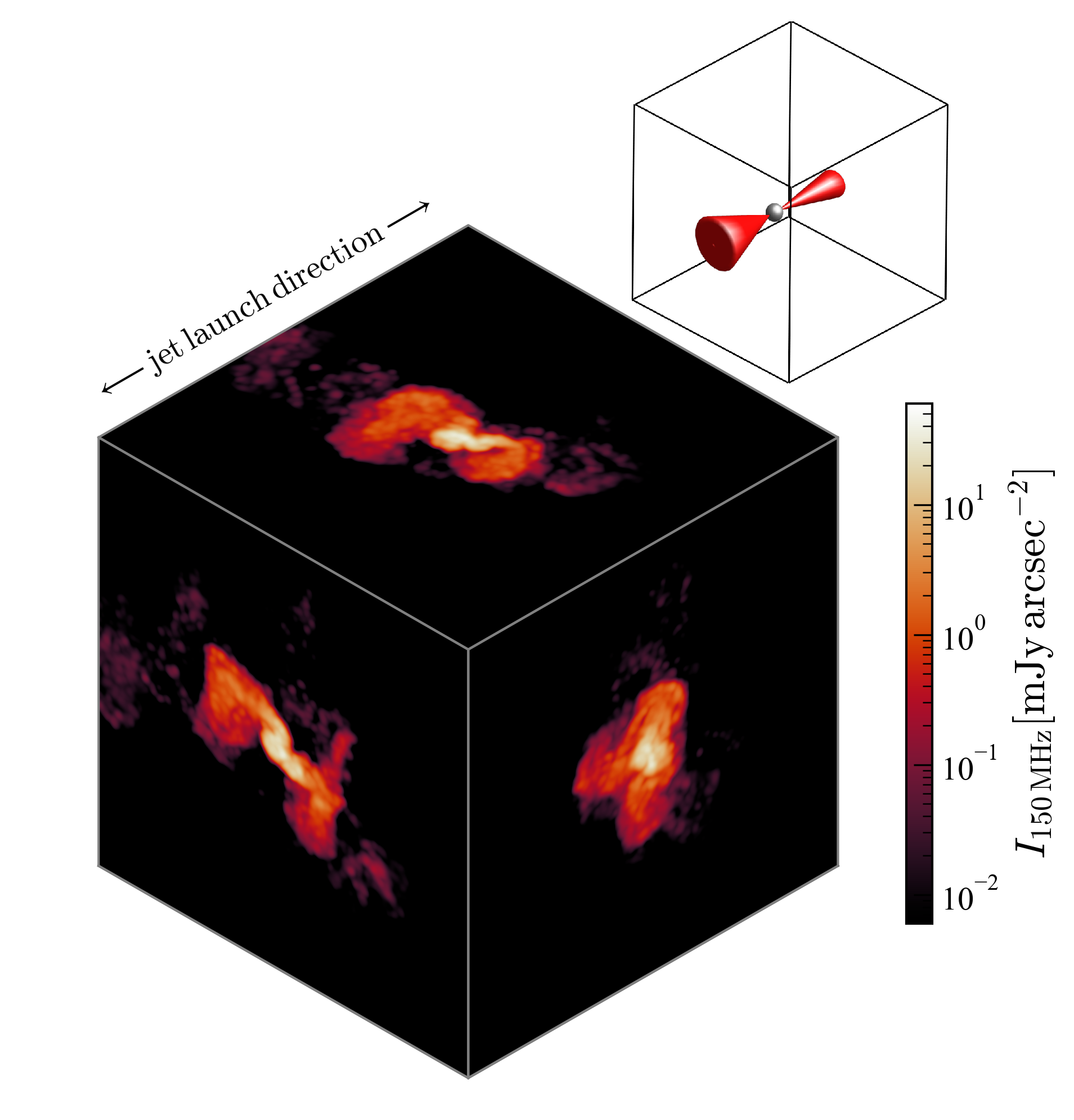}
	\caption{Synchrotron intensity maps projected along three different axes at $150$~MHz of a single snapshot at $t = 1.46$~Gyr (see \href{https://youtu.be/ADTj6sDtdZg}{movie}), and smoothed with a 2D Gaussian beam $\rm{FWHM} = 4$~kpc. The cartoon in the upper right illustrates the jet launching direction. The bottom left image in the cube is obtained by projecting along the jet launching axis, as is the case for the mock images shown in Fig.~\ref{fig:radio_maps}. This viewing direction corresponds to a blazar-like configuration, despite which we observe clear jet-inflated lobes. This demonstrates the deflection experienced by the jets on kpc-scales, caused by cold gas structures forming in the cluster centre and accreted onto the SMBH.  
	}
	\label{fig:emission_3angles}
\end{figure}

\subsection{Radio mocks: a diversity of radio morphologies and phenomena}\label{sec:results:radio-mocks}
In Fig.~\ref{fig:radio_maps}, we show radio intensity maps at 150~MHz at different epochs chosen to demonstrate the variety of obtained morphologies. For comparison, we display the radio emission for the single-jet outburst presented in \citetalias{Jlassi2026a} in the bottom right. The latter emerges from a jet event with power $L_\rmn{jet} = 3 \times 10^{45}$~erg~s$^{-1}$ active for 50~Myr, shown here at 55~Myr. We note that the radio images obtained for the self-regulated simulation are all projected along the axis parallel to the jet launching axis whereas that of the single outburst is projected perpendicular to the jet launching axis, for the purpose of comparing morphologies. We make use of the sensitivity reached in the LOFAR Two-metre Sky Survey (LoTSS) of $\sigma_{\rm{LoTSS}} = 100 \, \mu \rm{Jy / beam}$ \citep{Shimwell2017a} to plot contours at $\sigma_{\rm{LoTSS}} \times [3, 100, 800]$. We place our simulated object at the redshift of Perseus, $z = 0.0183$, which corresponds to an angular scale of $0.38 \, \rm{kpc} \, \rm{arcsec}^{-1}$. We adopt a beam size with full-width at half-maximum (FWHM) $= 7\arcsec$\footnote{This is comparable to the angular resolution in LoTSS as well as the highest resolution observations of the Perseus cluster \citep{Groeneveld2026} at this frequency.} applied as a symmetric 2D Gaussian kernel to smooth our radio intensity maps. 

The peak intensity at 150~MHz reaches levels of ${\approx}0.23  \, \rm{Jy} \, \rm{arcsec}^{-2}$, corresponding to ${\approx} 13 \,  \rm{Jy} \, \rm{beam}^{-1}$ for our smoothing beam of $7\arcsec\times7\arcsec$. This is on par with observations of radio galaxies with clear AGN lobes in galaxy clusters such as M87 in the Virgo cluster \citep{DeGasperin2012} or Hydra A in Abell 1958 \citep{Wise2007}, with levels of $10$--$100 \,\rm{Jy} \, \rm{beam}^{-1}$ obtained with larger beams. On the other hand, observations of radio lobes in the Perseus cluster display lower surface brightness \citep{Gendron-Marsolais2020} in comparison with our peak value. However, the lobes observed in Perseus do not exhibit clear jet structures and likely correspond to a weaker or older jet outburst. Generally, without fine-tuning parameters or rescaling our electron spectra, we successfully reproduce observed levels of radio surface brightness. This is also demonstrated by the global radio luminosities obtained throughout our simulation, which we discuss in Sect.~\ref{sec:results:global-electron-spectra}.

Furthermore, the first contour shown in Fig.~\ref{fig:radio_maps} corresponding to $3 \sigma_{\rm{LoTSS}}$ reveals that aged lobes from previous outbursts, at distances reaching ${\approx} 200$~kpc, would in principle be observable by such a survey. However, real radio observations, in comparison to our synthetic ones, still face numerous challenges such as incomplete $u$--$v$ plane coverage, calibration uncertainties, high dynamic range observations and the challenge of completely removing central or background radio sources which may give rise to side lobes, etc. 


\begin{figure*}
	\centering
	\includegraphics[width=2\columnwidth]{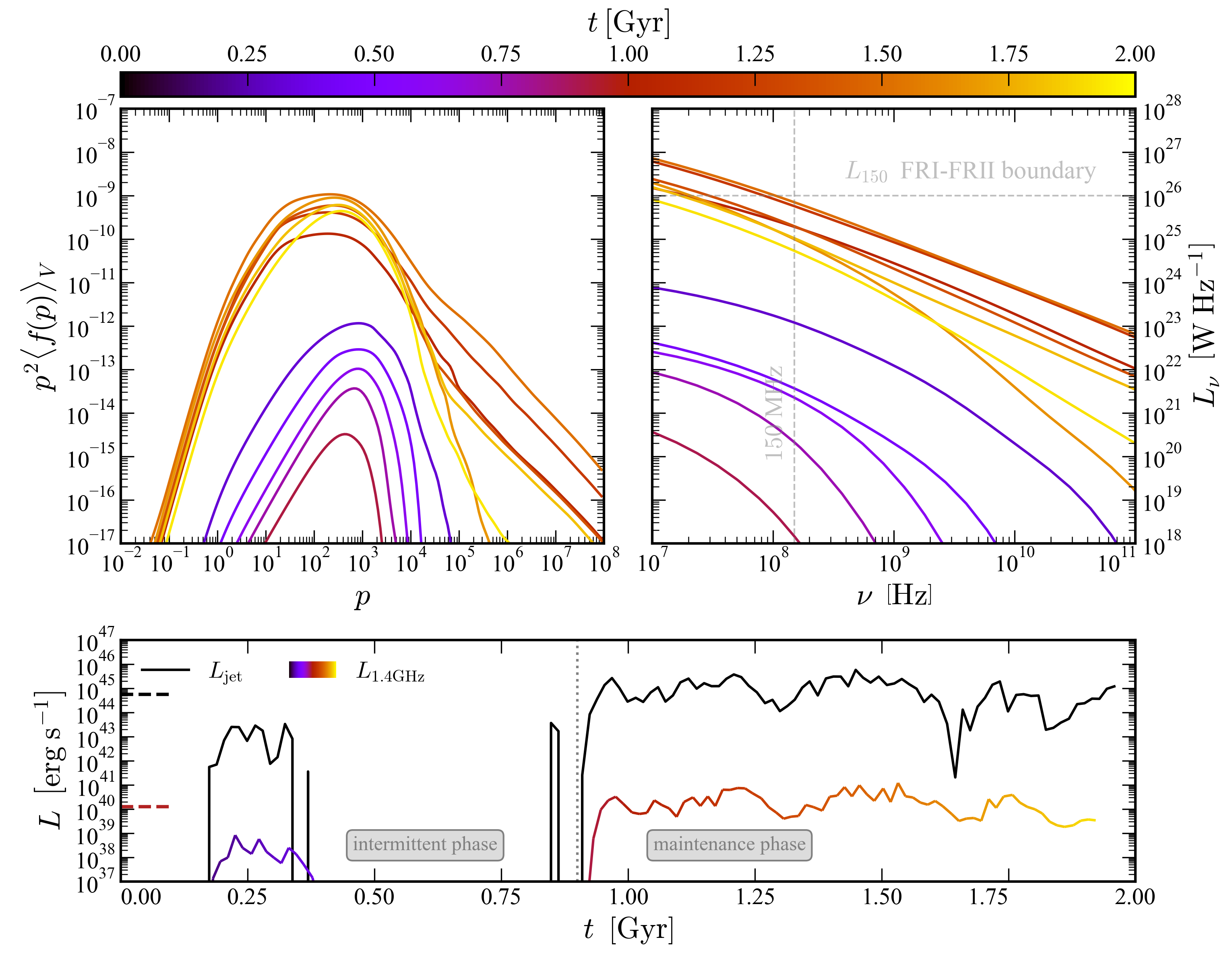}
	\caption{
		\textit{Bottom:} Time evolution of the jet luminosity (black line) and 1.4~GHz radio luminosity (coloured line). The black and red dashed lines indicate the mean jet and radio luminosities, respectively. The jets are initially in an `intermittent' phase with short, weak outbursts ($\leq 0.9$~Gyr) and transition into a `maintenance'-phase where the outbursts are powerful and almost continuous ($> 0.9$~Gyr).
		\textit{Top left:} Time evolution of the volume-weighted non-thermal electron spectra. In the first Gyr, the sparse jet outbursts lead to a strongly-cooled CR electron spectrum as a result of Coulomb and radiative losses. Once the AGN jets enter a maintenance phase with almost continuous outbursts, the high momentum part of the spectrum exhibits a steady-state slope as new electrons are accelerated.
		\textit{Top right:} Time evolution of the total radio luminosity between 10~MHz and 100~GHz computed from the electron spectra. The vertical dashed line indicates the $\nu = 150$~MHz frequency, and the horizontal dashed line the FRI-FRII radio power divide observed at this frequency \citep{Mingo2019}. The spectral evolution over time is analogous to that of the CR electron spectrum.
	}
	\label{fig:cre_spectrum} 
\end{figure*}

Throughout 2~Gyr of evolution, jet radio plasma reaches extents up to 400~kpc with diverse morphologies, with some clearly indicating the presence of jets, and others showing remnant plasma with complex morphologies. Images C, D, E, F resemble FRI radio galaxies, with plume-like, disturbed and bent jet lobes. In fact, applying the traditional definition used in classifying FRI and FRII sources \citep{Fanaroff1974} results in 97\% of our snapshots being identified as FRI sources in the AGN's maintenance phase (see Sect.~\ref{sec:results:global-electron-spectra}). However, this classification has limitations when applied to unusual and complex morphologies departing from ideal, polar lobes. We describe our procedure and display such edge cases in Appendix~\ref{sec:appendix:FR-classification}. Amongst these morphologies, we find a radio image with a single lobe in panel A, restarted radio galaxies in panels C, D, E and F,  i.e. evidence of episodic activity where an older pair of radio lobes from previous jet activity is observed simultaneously with a pair of inner lobes from ongoing activity \citep[e.g.][]{Schoenmakers2000, Jamrozy2007}. The single outburst presented in \citetalias{Jlassi2026a} and shown in panel I at 5~Myr post jet activity displays the typical, two-sided polar lobes observed in both FRI and FRIIs \citep{Capetti2017, Capetti2017a}. In contrast to this, images A, B and H exhibit radio plasma that is difficult to directly identify as typical pairs of jet lobes. Looking at the corresponding spectral index maps for these images (see Appendix~\ref{sec:appendix:spectral-index-maps}) reveals that the emission in B and G display low spectral indices of $\alpha^{1.4\,\rmn{GHz}}_{150\,\rmn{MHz}} \approx 0.6$ close to the SMBH, indicative of freshly accelerated plasma. Panel H, on the other hand, exhibits spectral indices larger than unity, and corresponds to radio plasma in a remnant phase, when the bright radio core and jets disappear but lobes continue to radiate.

As previously mentioned, the mock radio maps from our self-regulated AGN jets shown in Fig.~\ref{fig:radio_maps} were obtained by projecting the radio emission from CR electrons along the jet launching direction. To understand the implications of this, we select one epoch, $t = 1.46$~Gyr, and present its three-dimensional radio morphology. The corresponding radio intensity maps are presented in Fig.~\ref{fig:emission_3angles}, where each face of the cube corresponds to a different projection direction.

These three projections confirm the strong deflection experienced by the jets, causing the radio emitting plasma to spread into nearly all directions around the SMBH. Moreover, projecting along the jet launching direction corresponds to a blazar-like configuration where we are observing the jet `down the barrel'. All radio images shown in Fig.~\ref{fig:radio_maps}, as well as the left panel of Fig.~\ref{fig:emission_3angles} are hence blazar configurations, despite the complex lobe morphologies observed. In fact, while the radio image obtained in the blazar configuration clearly resembles a radio galaxy, projecting along a different direction yields a radio galaxy with what appears to be three lobes (rightmost side of the cube). Projecting along the final direction (upper side of the cube) helps disentangle this three-lobe configuration into a radio galaxy with strongly deviated jets, following an almost helical structure (we emphasize that we neither adopted a jet opening angle nor any precession). 

As discussed in Sect.~\ref{sec:results:jet-deflection} these complex and disturbed morphologies are due to light jets interacting with cold gas structures in the background gas, which additionally feed the SMBH through cold accretion (see Sect.~\ref{sec:AGN-feedback}). Indeed, both the self-regulated AGN jet simulation presented in this work and the single jet outburst presented in \citetalias{Jlassi2026a} employ identical initial conditions and jet density. The interplay between the radiatively cooling background gas and accretion driven jet activity is therefore responsible for the morphological differences observed between these two setups, which emphasises the need to model self-regulated jets in order to produce realistic radio galaxies in simulations.

\section{How AGN jet activity influences global electron and radio spectra}\label{sec:results:global-electron-spectra}

\subsection{Two evolutionary phases of jet activity}

Having explored the morphological characteristics of our AGN jets, we now present global CR electron and radio spectra responsible for this emission. Figure~\ref{fig:cre_spectrum} shows the time evolution of the jet and radio powers, the corresponding global CR electron spectrum (left) and the resulting radio spectrum (right) throughout 2 Gyr of evolution. Electron spectra are volume-weighted while radio spectra are obtained by integrating over the intensity of all tracer particles within a box of side 340~kpc centred on the SMBH.

Focusing first on the bottom panel, the temporal evolution of the jet luminosity clearly exhibits two phases. The first Gyr is an `intermittent' phase, where the jet outbursts are short and weak (${<} 10^{44} \, \rm{erg}\, \rm{s}^{-1}$). After nearly 1~Gyr, the AGN begins a more active phase owing to the cold gas mass accreted onto the SMBH and feeding the jet (see Fig.~\ref{fig:bh_diagnostics} for the temporal evolution of accretion and SMBH-related quantities). In this second period, which we call the `maintenance' phase, the AGN is active with longer outbursts of higher average jet power (${>} 10^{44} \, \rm{erg}\, \rm{s}^{-1}$). In this maintenance phase, there are nonetheless periods where the jets nearly switch off, e.g. $t \approx 1.6$~Gyr due to a decrease in the accreted cold gas mass. Observationally, this epoch would correspond to a case of a restarted radio galaxy, where the decrease in jet luminosity would be observed as jets switching off, as is the case in our simulated radio maps (see \href{https://youtu.be/HJEeYl__gPU}{movie}).

We show the average jet and radio luminosities as dashed black and red lines, respectively. Our simulations yield an average jet power of $5.7 \times 10^{44} \, \rmn{erg} \, \rmn{s}^{-1}$ and an average specific 1.4~GHz radio luminosity of $1.2 \times 10^{40} \, \rmn{erg} \, \rmn{s}^{-1}$. The radio power, smaller by roughly four orders of magnitude, traces the jet luminosity but remains on a floor value throughout the maintenance phase even when the instantaneous jet luminosity drops abruptly, as is the case around $1.6$~Gyr. This can be explained by the variability timescale of jet activity being shorter than the cooling time scale of CR electrons which are progressively accelerated (rather than injected at a unique event in a jet outburst). Electrons hence still emit radio emission despite the decrease in instantaneous jet power. In future work, we will investigate the link between jet and radio luminosities.

Overall, our simulations yield radio luminosities below $10^{26} \, \rm{W} \, \rm{Hz}^{-1}$ at $150$~MHz, in agreement with observed luminosities of FRI sources \citep{Fanaroff1974, Ledlow1996, Mingo2019}, which have lower luminosities in comparison to FRII sources\footnote{This break, however, is only an indicator of where FRI and FRII sources are expected to lie on average; there is a large overlap in the luminosity associated with objects of these two morphologies \citep{Mingo2019}.}. This result is consistent with the radio intensity levels obtained in our maps (Fig.~\ref{fig:radio_maps}), both of which are in line with observations.

\subsection{Intermittent phase}
We now focus on the top panels of Fig.~\ref{fig:cre_spectrum}. In the initial, intermittent phase (shown in purple to magenta-coloured lines), the electron spectrum exhibits properties of a typical freely-cooling spectrum: a convex spectrum shaped by Coulomb losses at low momenta, and synchrotron and inverse-Compton losses at high momenta. The spectrum reduces in normalization and narrows with time as electrons are progressively cooled after the initial outbursts. Comparing this spectral evolution to that without adiabatic changes and mixing effects included (described in Sect.~\ref{sec:adiabatic-effects} and in extensive detail in \citetalias{Jlassi2026a}) confirms that CR electrons experience dilution, which explains this decrease in the normalisation. The overall freely-cooling behaviour of the spectrum is particularly noticeable in this intermittent phase where jet outbursts are weak and short-lived because they form significantly smaller lobes than in the maintenance phase. These lobes do not propagate further than 50~kpc and are easily disrupted through mixing with the ICM. The temporal evolution of the radio spectrum mirrors this freely-cooling behaviour, with synchrotron losses being faster at higher frequencies. As a consequence, the radio luminosity at 150~MHz (cf. vertical dashed line) decreases by nearly five orders of magnitude in about $0.5$~Gyr, and is two orders of magnitude lower than the luminosity in the maintenance phase (shown in red to yellow-coloured lines). For these reasons, we choose to show radio morphologies from the maintenance phase in Fig.~\ref{fig:radio_maps}. The morphology observed in the initial intermittent phase are instead compact and of very low luminosity, and could be likened to unresolved, compact radio galaxies, also referred to as FR0 sources \citep{Sadler2014, Baldi2015, Baldi2019, Vardoulaki2021}.

\begin{figure*}[ht]
	\centering
	\includegraphics[width=2.\columnwidth]{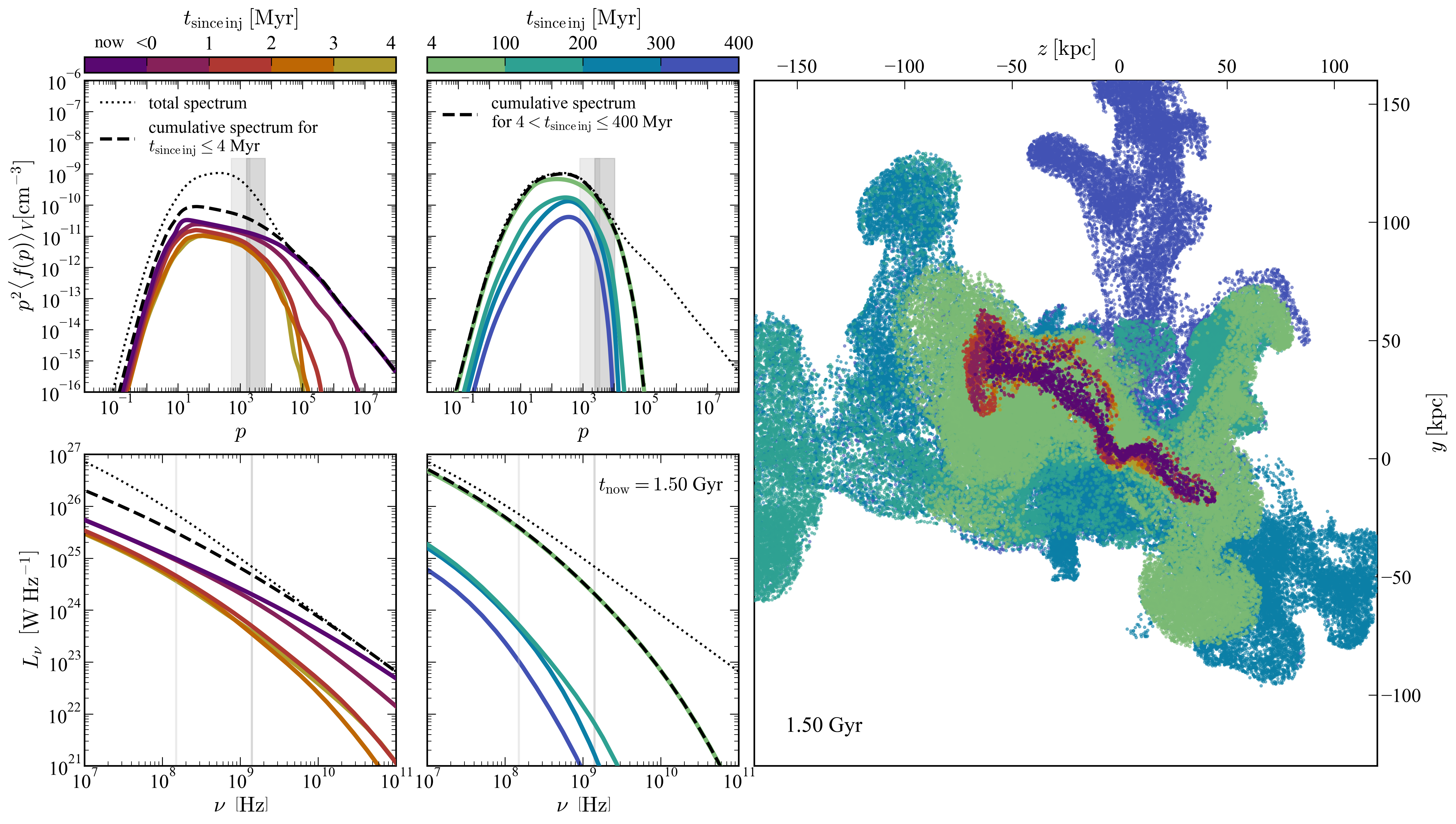}
	\caption{Spatial and spectral distributions of different electron ages, defined as the time since last injection $t_{\rmn{since \, inj}}$, at $t = 1.5$~Gyr. We differentiate between young (age $\leq 4$~Myr) and old ($4 <$ age $\leq 400$~Myr) CR electrons on the left and right, respectively. Electrons undergoing acceleration at the current time correspond to $t_{\rmn{since\, inj}} = \rm{now}$.
		\textit{Right:} Spatial distributions of CR electron tracer particles, colour-coded by their age, i.e. their corresponding time since injection. 
		\textit{Left, top row: } CR electron spectra colour-coded by the time since injection. Using the 10-90th percentiles of the magnetic field strength distribution, we plot the electron momentum contributing to the 150~MHz and 1.4~GHz emissions in light and dark grey, respectively. We show the global spectrum at the snapshot time as a dotted black line, and the cumulative spectra composed of the electron populations showed in each panel using a dashed black line.
		\textit{Left, bottom row:} Corresponding synchrotron radio spectra. High momenta ($p > 10^5$) are dominated by electrons undergoing acceleration. Low and mid-momenta ($p < 10^4$) are dominated by old electrons. Both young and old CR electrons contribute to the intermediate momentum range $10^3 < p < 10^4$, which is precisely the range which contributes to emission in the 150~MHz -- 1.4~GHz band.
        }
    \label{fig:spectra_binned_by_time_since_inj}
\end{figure*}

\subsection{Maintenance phase}
In the second Gyr of evolution where the AGN is in a maintenance mode, the global CR electron spectral shape resembles a freely-cooling spectrum at low and mid-momenta ($p < 10^4$), while high-momenta display a steady-state slope of $\alpha_{\rm{inj}} + 1 = 3.2$ almost throughout. This steady-state slope indicates that ongoing acceleration is competing with radiative losses at these momenta. Despite this, the spectral normalisation at high momenta varies by a factor of 100. As described in Sect.~\ref{sec:cre-acceleration}, 0.1\% of the jet energy is injected into electrons. Assuming all radio emitting plasma experiences similar levels of dilution (in turn decreasing the spectral normalisation by the same amount), the change in normalisation at mid-momenta is therefore a consequence of the varying jet luminosity. The radio spectrum mirrors the electron spectrum and generally exhibits a power-law behaviour. A short interruption in AGN activity and hence a lack of newly-injected CR electrons e.g. at around $1.6$~Gyr, is reflected in the CR electron spectrum, which again displays a freely-cooling shape due to dominating radiative losses. This is also visible in the radio spectrum, which deviates from a power-law behaviour at this time. This indicates that the high-momentum spectrum ($p > 10^6$) retains almost no memory of recent injection events due to fast cooling losses. To understand this, we analyse how populations of different ages contribute to the global spectrum at a single epoch in the following section.

\section{Disentangling the impact of electrons of different ages on the resulting emission}\label{sec:results:contributions-different-ages}

We distinguish between populations of CR electrons accelerated at different stages in Fig.~\ref{fig:spectra_binned_by_time_since_inj} and focus on a single snapshot at $t=1.5$~Gyr, during which the jet is active. Specifically, we distinguish between electron populations accelerated in the last 4~Myr in the first column, and electrons accelerated between 4 and 400~Myr in the second column, and show their physical location in the third column. We make use of the quantity $t_{\rmn{since\, inj}}$, defined as the time elapsed since the last CR electron acceleration event\footnote{More specifically, this corresponds to the last integration timestep in \textsc{Crest} where CR electron energy was injected} for an individual tracer particle (we describe the criteria for acceleration in Sect.~\ref{sec:CRes}). Because the time since the last injection $t_{\rmn{since\, inj}}$ can be defined as the electron age, we use this terminology in the following description\footnote{In \citetalias{Jlassi2026a}, two age definitions were adopted depending on the electron spectral range of interest. For the purpose of this analysis, which concerns Gyr timescales, the 10~Myr progressive injection timescale introduces a negligible distinction between the two definitions, and we therefore do not differentiate between them here for simplicity.}, and where unspecified, we generally refer to young and old electrons as electrons with ages $\leq 4$~Myr (first column) and $4~\rmn{Myr} < t_{\rmn{since\, inj}} \leq 400$~Myr (second column), respectively.

\subsection{CR electron age distributions: electron spectra}\label{sec:results:age-distributions-cre-spectra}

Focusing first on CR electron spectra, electrons still undergoing acceleration (purple line, $t_{\rmn{since\, inj}} = \rm{now}$) contribute most strongly to the high momentum spectrum from $p > 2 \times 10^5$, and trace the ongoing jet outburst in the map. The cumulative spectrum for all electrons younger than 4~Myr (black dashed line) also reveals that these electrons undergoing acceleration are solely responsible for the steady-state slope observed at high momenta in the global electron spectrum (black dotted line). Electrons no longer experiencing acceleration, but younger than 1~Myr have already cooled away at $p>10^7$ (magenta line), and are tracing a lobe at $(z, y) \approx (-60, -40)$~kpc. This trend continues for other young electrons, e.g. with ages between 1 and 4 Myr, as strong radiative losses carry on cooling these populations at high momenta. The spectral normalisation at $10^1 < p < 10^4$ varies across these populations. This decrease in normalization could be due to dilution and compression occurring simultaneously, shifting the spectrum right and downwards. It could also originate from the progressive and exponentially decreasing source function accelerating CRs in our simulations, according to which the spectral normalisation is determined by the maximum energy density injected into electrons. Therefore, as the jet power determines the energy accelerated into CRs, variations in the spectral normalisation of different electron populations are expected due to the fluctuating jet power. For a more detailed description of how our acceleration algorithms affect CR electrons spectra in a single-jet outburst, we refer the reader to \citetalias{Jlassi2026a}.

Shifting our attention to older electrons (middle column), the low- and mid-momentum electron spectral range ($p < 10^5$) in the global spectrum is dominated by contributions from electrons aged between 4 and 400 Myr. These display a typical freely-cooling behaviour of progressively losing energy through Coulomb losses and radiative losses, which narrows their spectral distribution. The CR electron spectral range $3 \times 10^4 < p < 10^5$ stems mostly from electrons with ages ranging from 4 to 100~Myr (as evidenced by the cumulative spectrum for old electrons), which trace the lobes of a previous jet outburst, reaching a horizontal extent of ${\sim} 200$~kpc ($-100~\rmn{kpc} < z < 100~\rmn{kpc}$) and vertical extent of ${\sim} 150$~kpc ($-75~\rmn{kpc} < y < 75~\rmn{kpc}$). Electrons older than 100~Myr still contribute strongly to the global spectrum due to the longer cooling times of electrons at $p < 3 \times 10^4$, and correspond to older jet outbursts. These old electrons have a heterogeneous distribution, reaching lateral extents greater than $250$~kpc and are located in almost all directions, emphasising the complex interaction between low-density jets and a turbulent environment forming cold gas structures (see Sect.~\ref{sec:results:self-regulation-radio-mocks}).

In conclusion, while old electrons determine the normalisation of the global electron spectrum at $p < 10^4$, only electrons undergoing acceleration are responsible for the high momentum, steady-state spectral shape. In the following, we examine the intermediate spectral range at momenta $10^3 < p < 10^4$, which correspond to the electron momenta that typically emit radio synchrotron emission in the range from 150 MHz to 1.4 GHz.

\begin{figure*}[ht]
	\centering
	\includegraphics[width=2\columnwidth, trim={2.2cm 0 3.2cm 0},clip]{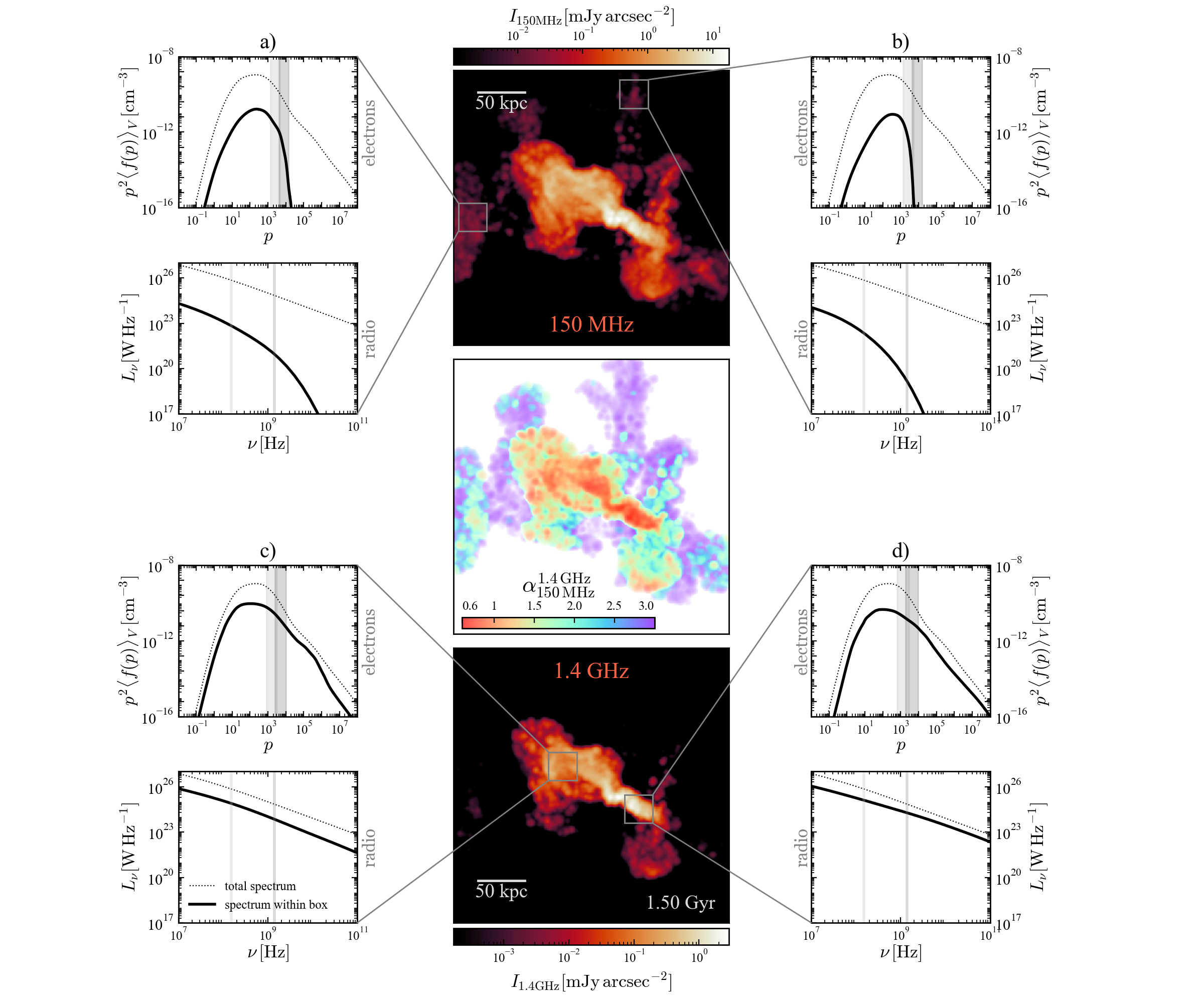}
	\caption{Multi-panel view of the AGN jet outburst at $t = 1.5$~Gyr (same time as Fig.~\ref{fig:spectra_binned_by_time_since_inj}). 
		\textit{Central column:} From top to bottom, we show the 150~MHz radio intensity, the spectral index map at 150~MHz -- 1.4~GHz, and the 1.4~GHz intensity. The radio mocks are smoothed with a Gaussian beam with $\rm{FHWM} = 4$~kpc, from which the spectral index map is computed. The multi-epoch radio intensity at these two frequencies is shown in a \href{https://youtu.be/HJEeYl__gPU}{movie}.
		\textit{Left and right columns:} CR electron spectra and corresponding radio synchrotron spectra in four different regions of interest. Each box has dimensions $30 \times 30 \, \rmn{kpc}^2$, and the same depth as the radio images. The 10--90th percentile magnetic field strengths within these boxes is used to show the momenta contributing to the 150~MHz and 1.4~GHz emission frequencies in light and dark gray, respectively.
		The similar features observed in the jet lobes at low and high frequencies are explained by the contributing momenta highlighting both old and young electrons (see Fig.~\ref{fig:spectra_binned_by_time_since_inj}). Regions a) and b) are composed of old electron populations whose cut-off momentum $p\approx10^4$ is in the band of contributing momenta, which is why these electrons are visible at 150~MHz, but not at 1.4~GHz. Regions c) and d) include younger electrons in comparison to regions a) and b), indicated by the power-law shape in region c) and d).
	}
	\label{fig:spectra_at_diff_locations}
\end{figure*}

\subsection{CR electron age distributions: radio synchrotron spectra}\label{sec:results:age-distributions-radio-spectra}

In the resulting synchrotron emission spectra (lower panels in Fig.~\ref{fig:spectra_binned_by_time_since_inj}), electrons younger than 4~Myr old dominate the global radio spectrum from ${\approx} 5$~GHz and at higher frequencies, as indicated by their cumulative spectrum (black dashed line). On the other hand, up to ${\approx} 200$~MHz, it is electrons aged between 4 and 100~Myr (light green line) which dominate the cumulative spectrum, with electrons older than 100~Myr not contributing at all. The cumulative spectra of young and old electrons strongly overlap in the intermediate spectral range $ 200 \, \rm{MHz} < \nu < 5 \, \rm{GHz}$. To understand this, we connect electron and synchrotron spectra using Eq.~\eqref{eq:p-contributing}, which allows us to determine which momentum $p$ contributes to the emission at a given frequency $\nu_{\rmn{sync}}$ given the magnetic field strength encountered by CR electrons. Histograms of the magnetic field strengths experienced by these CR electron populations are shown in Appendix~\ref{sec:appendix:Bfields-histograms} and range from 0.2 to 100~$\mu \rm{G}$ with most tracers located in fields of $2-40 \, \mu \rmn{G}$ (10th to 90th percentiles). These magnetic field strengths are in accordance with estimates obtained from observations \citep[see][for a review]{Carilli2002} using equipartition arguments \citep[e.g.][]{Birzan2008, DeGasperin2012} in AGN lobes and Faraday rotation measures \citep[e.g.][]{Vogt2005, Kuchar2011, Osinga2025} in the ICM.

We show the momenta contributing to the 150~MHz and 1.4~GHz emission in light and dark grey, respectively, in the top panels of Fig.~\ref{fig:spectra_binned_by_time_since_inj}, and indicate these frequencies in the synchrotron spectra using vertical lines of the same colours\footnote{The range of momenta shown corresponds to the minimum 10th percentile and maximum 90th percentile magnetic field strengths of the electron populations shown in each panel.}. This allows us to directly map electron spectral range to synchrotron frequency. Due to the comparable magnetic field strengths encountered by both young and old electrons (see Appendix~\ref{sec:appendix:Bfields-histograms}), momenta contributing to the emission at $150\,\rmn{MHz}< \nu < 1.4$~GHz generally fall in the range $5 \times 10^2 < p < 10^4$ for both young and old populations. This range is precisely where the global spectrum transitions between its freely-cooling, convex shape to its power-law behaviour, as previously mentioned. Indeed, due to cooling times longer than ${\sim} 200$~Myr at these momenta, electrons less than $100$~Myr old are able to contribute to the synchrotron emission in the $200 \, \rm{MHz} < \nu < 5 \, \rm{GHz}$ range. Nonetheless, in the band of momenta responsible for the 150~MHz emission, the spectrum corresponding to $4\,{\rm Myr} < \rmn{age} \leq 100$~Myr has a higher normalisation $p^2\langle f(p) \rangle_{V} \approx 2 \times 10^{-10}\,$cm$^{-3}$ in comparison to young electrons, and therefore dominate the emission spectrum at this frequency. The picture becomes more complex at 1.4~GHz, as both young and old electrons have equivalent spectral normalisations $p^2\langle f(p) \rangle_{V} \approx 3 \times 10^{-11}\,$cm$^{-3}$, albeit  with different spectral slopes. Young electrons have flatter spectra, which explains their larger contribution to the global radio spectrum at 1.4~GHz.

This effect -- whereby electrons of different ages contribute to the $150\,\rmn{MHz} <\nu <1.4\,\rmn{GHz}$ emission range -- explains why emission maps at similar spatial resolutions but different frequencies can look remarkably similar, as is the case for M87, the central galaxy in the Virgo cluster \citep{DeGasperin2012, DeGasperin2025}, which displays comparable features at 54, 144 and 1284~MHz \citep[see also][where the FRI source, Hydra A, appears similar at 74 and 330~MHz]{Lane2004}. To further illustrate this effect, we present 150~MHz and 1.4~GHz radio intensity maps in Fig.~\ref{fig:spectra_at_diff_locations} at the same epoch. Although we obtain generally similar maps at these two frequencies, the lower frequency map (middle column, top panel) reveals older radio emitting plasma in the form of lobes. The regions of high resemblance between the two maps correspond to electrons younger than 100~Myr old in Fig.~\ref{fig:spectra_binned_by_time_since_inj}, all of which have non-negligible CR electron spectra in the range of momenta contributing to the 1.4~GHz emission, as explained above.

Moreover, connecting electron and radio spectra using the encountered magnetic field strengths explains why extending observations to frequencies lower than 150~MHz does not always uncover significantly larger extents in emission. For example, the respective electron spectra in Fig.~\ref{fig:spectra_binned_by_time_since_inj} (left, top row) for populations with ages between 100-200 (teal colour) and 200-300~Myr (blue colour) look remarkably similar at $p \approx 10^3$, which contributes to the 150~MHz emission. As a consequence of their equivalent electron spectra, their radio spectral shape (left, bottom row) below 150~MHz is almost identical despite their different ages. While the luminosity observed at a frequency of 50~MHz would be greater than at 150~MHz, electrons with ages $100~\rmn{Myr} < t_{\rmn{since~ inj}} \leq 200$~Myr would not appear brighter than electrons with ages $200~\rmn{Myr} < t_{\rmn{since~ inj}} \leq 300$~Myr. Nonetheless, the general argument according to which lower frequencies probe older electrons still stands in terms of relative luminosities. That is to say, for the oldest electrons with ages $300 ~\rmn{Myr}< t_{\rmn{since~ inj}} \leq 400$~Myr and assuming equal observational sensitivities at 50~MHz and 150~MHz, observations at lower frequencies increases their likelihood of detection, simply due to their greater luminosity at lower frequencies.

\subsection{Connecting age distributions to local spectra}\label{sec:results:connecting-ages-to-local-spectra}

Figure~\ref{fig:spectra_at_diff_locations} highlights the localised spectral information encoded in our simulations. Four different regions are shown, each with its corresponding local CR electron and radio spectrum. The range of momenta contributing to the 150~MHz--1.4~GHz emission lies around $10^3 < p < 10^4$ and is uniform across all regions.

Region d) is centred on the jet, and is almost identical to the cumulative spectrum for electrons with $t_{\rmn{since\,inj}}\leq 4$~Myr (upper left panel of Fig.~\ref{fig:spectra_binned_by_time_since_inj}), which indicates that the youngest electrons -- those still undergoing acceleration -- are contributing in this region. The spectral index map reflects this with a smooth distribution of $\alpha_{\nu} = 0.6$ along the jet barrel. Region c) highlights a recently inflated jet lobe with a spectrum exhibiting a power-law behaviour at high momenta, indicative of ongoing acceleration. However, the CR electron spectrum in this region has similar features to the global spectrum (black dotted line in Fig.~\ref{fig:spectra_binned_by_time_since_inj}), notably the freely-cooling shape at $10^{-1} < p < 10^3$. Indeed, despite the ongoing acceleration of CR electrons, the spectral index distribution in this region includes values higher than $\alpha_{\nu} \approx 0.6$. This region appears to contain the entire range of CR electron populations accelerated at different jet outbursts, which explains why the spectrum in this region has similar features to the global spectrum.

Regions a) and b) emphasise old lobes, i.e. remnant plasma with age ranges of 100--200~Myr and 300--400~Myr, respectively. Despite having different ages, their spectra have similar, freely-cooling shapes. The larger age of electrons in region b) (cf. Fig.~\ref{fig:spectra_binned_by_time_since_inj}) explains the spectral cut-off at a lower momentum in comparison to that in region a), due to CR electrons having cooled for a longer duration. Specifically, this cut-off is at the edge of the band of momenta contributing to the 1.4~GHz, which explains the lack of visible emission in the 1.4~GHz image in this region. On the other hand, in region a), electrons in the 1.4~GHz momentum range have not cooled completely, which explains why they are slightly visible in the 1.4~GHz image.

\section{Conclusions}
In this study, we extend the self-regulated Perseus-like cool-core cluster presented in \citetalias{Ehlert2023} by accelerating and evolving CR proton and electron populations in the AGN jets. The CR protons are modelled as a relativistic fluid in MHD, while CR electrons are evolved in post-processing along Lagrangian tracer particles using \textsc{Crest} \citep{Winner2019} to evolve CR electron spectra. We use \textsc{Crayon+} to calculate the radio synchrotron emission \citep{Werhahn2021c}. The AGN jet luminosity is self-consistently determined by cold accretion onto the central SMBH and hence fluctuates throughout the simulation duration of 2~Gyr. In this work, we present spatially and temporally evolving CR electron spectra and the resulting mock radio continuum observations. We summarize our findings as follows:

\begin{itemize}
	\item Self-regulated AGN jets produce complex and `disturbed' FRI morphologies (Figs.~\ref{fig:radio_maps} and \ref{fig:emission_3angles}) which depart from typical polar lobe structures prevalent in numerical studies of AGN jets. Indeed, comparing these complex morphologies to the polar lobe structures obtained in our previous work \citepalias{Jlassi2026a} which uses identical initial conditions and algorithms but crucially, a single jet outburst of fixed power, demonstrates that accretion-modulated AGN jet power (Fig.~\ref{fig:cre_spectrum}) plays a major role in producing realistic morphologies (Fig.~\ref{fig:radio_maps}). Moreover, the variability in jet power produces multiple generations of lobes, as well as morphologies that match restarted and remnant radio galaxies.

	\item All mock radio observations presented are viewed in a blazar-configuration -- with the line of sight parallel to the jet launch direction -- and yield intricate lobe and plume features despite this, emphasizing the strong deflection of the jets on kpc scales (Figs.~\ref{fig:radio_maps} and \ref{fig:emission_3angles}). This deflection is due to our simulated light jets ($\rho_{\rmn{ICM}} / \rho_{\rmn{jet}} \sim 3 \times 10^3-10^4$), which carry a low momentum density and are hence easily re-oriented by thermally unstable cold and dense gas structures, the latter which also feed the central SMBH through cold accretion and power the jet \citepalias{Ehlert2023}. Thus, our simulations suggest that inferring the black hole spin and small-scale jet orientation from (FRI) radio lobe morphologies on kpc scales and above is, fundamentally, not possible.

    \item Our self-consistently evolved CR electron spectra allow us to account for a well-known observational phenomenon: radio observations of AGN jet lobes in cluster environments occasionally display similar spatial extents and morphological features across different frequencies (Fig.~\ref{fig:spectra_at_diff_locations}). This is explained by the range of electron momenta $10^3 < p < 10^4$ contributing to the 150~MHz--1.4~GHz frequency range in magnetic fields of $1$--$50 \, \mu \rmn{G}$ obtained in our cool-core environment. In this range of momenta, both young (a few Myr old) and old (hundreds of Myr old) electrons are able to contribute to the radio emission observed in the 150~MHz--1.4~GHz range (Fig.~\ref{fig:spectra_binned_by_time_since_inj}).

\end{itemize}
	
The modelling of temporally and spatially evolving CR electron spectra allows us to examine the link between CR electrons shaped by the plasma they are embedded in, and the resulting observed radio emission. Our simulation setup is ideally suited to study various radio phenomena relating to AGN jets in galaxy clusters. Moreover, on top of the radio diagnostics they provide, our AGN jets' ability to prevent cooling flows allows us to study the connection between jet power and observed radio emission, to gain insight into AGN feedback from the radio window. We will explore these topics in future work.

\begin{acknowledgements}
	The analysis of the simulations presented in this work were performed using the Python package \textsc{Paicos} \citep{Berlok2024}. LJ and CP  acknowledge support from the Deutsche Forschungsgemeinschaft (DFG, German Research Foundation) as part of the DFG Research Unit FOR5195 – project number 443220636. CP and JW acknowledge support by the European Research Council under ERC-AdG grant PICOGAL-101019746. RW acknowledges funding of a Leibniz Junior Research Group (project number J131/2022). PG gratefully acknowledges financial support from the European Research Council via the ERC Synergy Grant "ECOGAL" (project ID 855130).
\end{acknowledgements}

\section*{Data Availability}

The data underlying this article will be shared on reasonable request to the corresponding author.

\bibliographystyle{aa}
\bibliography{paper2.bib}

\begin{appendix}

\nolinenumbers

\section{Energetics and self-regulation of AGN jet feedback over 2~Gyr}\label{sec:appendix:self-regulation-total-energetics}

\begin{figure}
	\centering
	\includegraphics[width=0.9\columnwidth]{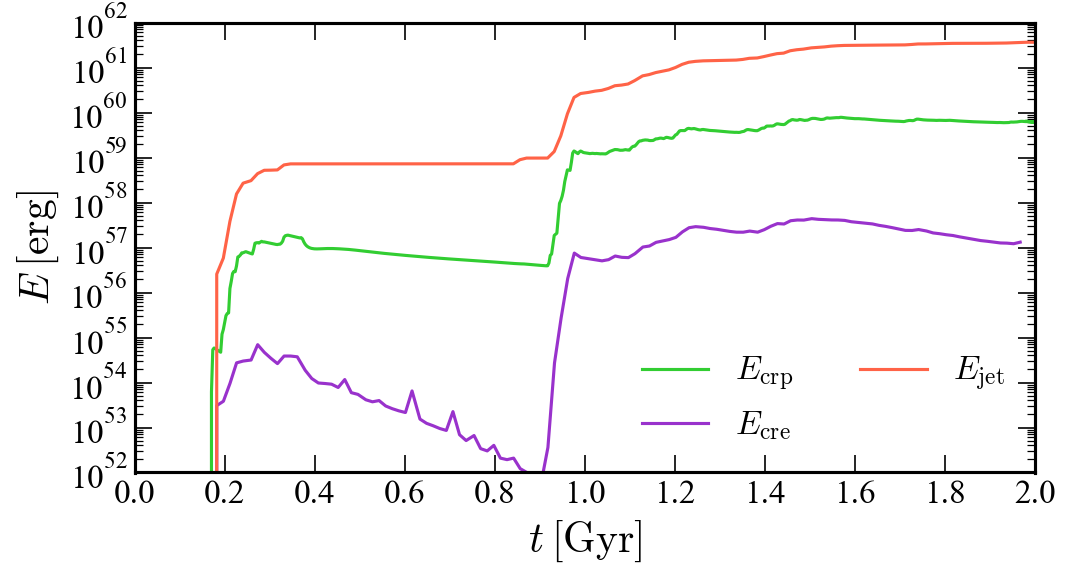}
	\caption{
		Time evolution of the cumulative jet, CR proton, and CR electron energies. Protons are transported through advection and diffusion, and cooled through Coulomb, hadronic, and Alfv\'en losses. The electrons are transported strictly through advection, and cooled through Coulomb and radiative (bremsstrahlung, synchrotron and IC) losses.
        The different loss processes experienced by CR protons and electrons explain the differently decreasing CR proton and electron energies over time.}
	\label{fig:energies}
\end{figure}

\begin{figure}[ht]
	\centering
	\includegraphics[width=1\columnwidth]{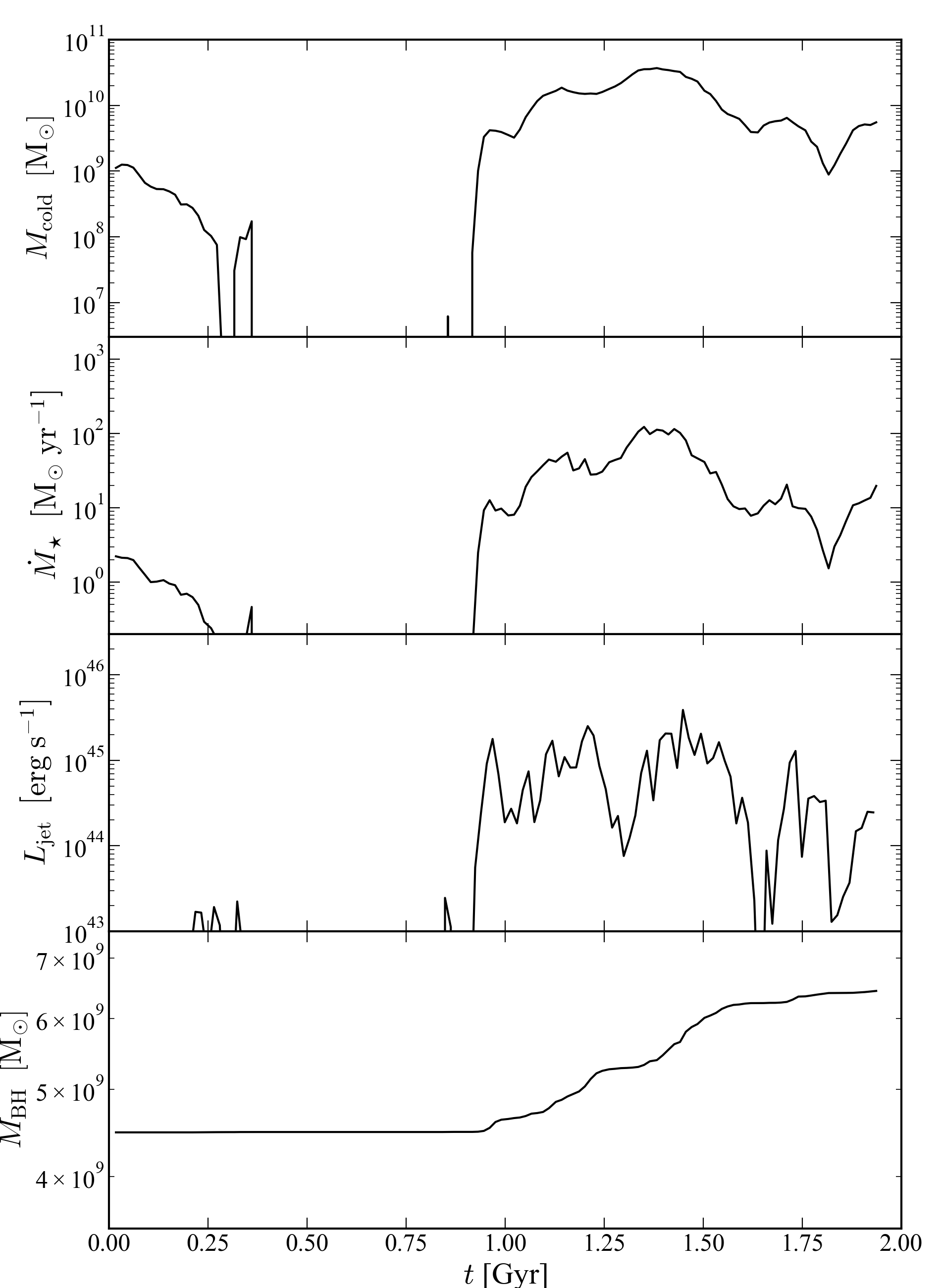}
	\caption{Time evolution of cold gas mass $M_{\rm{cold}}$ ($T < 10^6 \rmn{K}$), star formation rate $\dot{M}_{\rm{\star}}$, jet power $L_{\rm{jet}}$ and black hole mass $M_{\rm{BH}}$. The AGN jet model establishes self-regulation, most clear at $t > 0.9$~Gyr, in the maintenance phase where the jet is continuously on.}
	\label{fig:bh_diagnostics}
\end{figure}

\begin{figure}[ht]
	\centering
	\includegraphics[width=1\columnwidth]{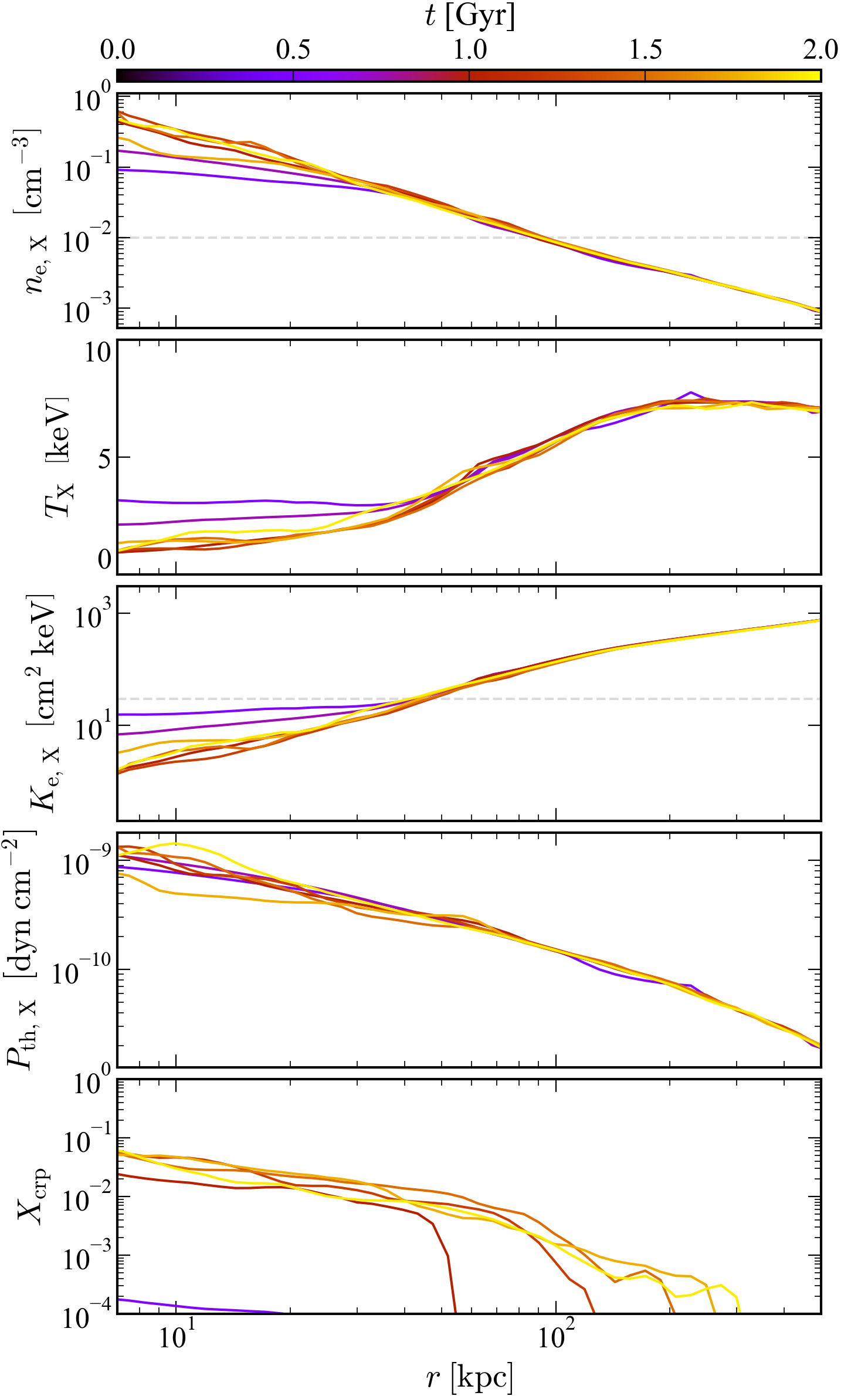}
	\caption{Radial profiles of the X-ray weighted electron number density $n_{\rmn{e, X}}$, temperature $T_\rmn{X}$, entropy $K_\rmn{e,X}$, thermal pressure $P_\rmn{th, X}$. The final panel shows the new addition of CR protons to the \citetalias{Ehlert2023} setup through the CRp-to-thermal pressure ratio $X_\rmn{crp}$, which uses volume-weighted pressures. Dashed grey lines indicate number density ($10^{-2} \, \rmn{cm}^{-3}$) and entropy ($30 \, \rmn{keV}\,\rmn{cm}^{2}$) criteria used to separate strong cool-core clusters from moderate to non cool-core clusters \citep{Cavagnolo2008, Hudson2010}.
    The thermodynamic profiles are consistent with observations and stable after 0.5~Gyr, highlighting the self-regulation achieved in the cool-core cluster.}
	\label{fig:radprof_thermodynamics}
\end{figure}

In the top panel of Fig.~\ref{fig:energies}, we show the evolution of the cumulative injected jet energy, and CR proton and electron energies. The initial short jet outbursts occurring around 0.25 Gyr (see Fig.~\ref{fig:bh_diagnostics}), are visible as an increase in energy for all components. These outbursts are followed by a quiescent period during which the CR proton and electron energies decrease, albeit the latter more drastically. This decrease is due to CR protons experiencing Coulomb, hadronic and Alfv\'enic losses, while electrons experience Coulomb and radiative losses as well as adiabatic and mixing losses. Although the overall CR electron energy decreases between $0.3$ and $0.9\,{\rm Gyr}$, peaks are observed, likely corresponding to discrete accretion events. In the second Gyr, where the AGN is in a maintenance phase (see Fig.~\ref{fig:bh_diagnostics}), the cumulative jet energy increases, allowing the total CR proton and electron energies to somewhat stabilise despite their strong losses. From 1.6~Gyr onwards, the average jet power decreases, leading the total CR electron energy to decrease overall. This can be explained by newly injected CR electrons carrying less energy than previous jet events, which are therefore not able to replenish the electron energy fast enough to counteract Coulomb and radiative losses. For a more detailed discussion on the acceleration algorithms and the recovered energy fractions, we point the reader to \citetalias{Jlassi2026a}, where we show the evolution of jet, CR proton and electron energies in a simpler setup where we simulate a single jet outburst.

We now present the self-regulating nature of our AGN jet feedback model. To do this, we present similar diagnostics as those presented in \citetalias{Ehlert2023}. In Fig.~\ref{fig:bh_diagnostics}, we show the evolution of properties related to the cooling processes, star formation and the accreting SMBH, namely the cold gas mass $M_{\rm{cold}}$ ($T < 10^6$~K), the star formation rate $\dot{M}_{\star}$, the jet power $L_{\rm{jet}}$, and the black hole mass $M_{\rm{BH}}$. The initial intermittent phase of AGN activity is due to cold accretion which fuels the AGN jets. 
The range of values recovered in our simulation align with those obtained in Fig.~6 of \citetalias{Ehlert2023}, albeit with slightly higher cold gas mass and hence higher star formation rates ($< 200 ~\rmn{M}_{\odot} ~\rm{yr}^{-1}$). The addition of CR protons in comparison to their work likely has an impact on the evolution of these quantities, which will be explored in future work. This novel CR proton component is presented in Fig.~\ref{fig:radprof_thermodynamics} in conjunction with radial profiles of X-ray weighted thermodynamic quantities characterizing cool-core clusters.  The X-ray weighting uses the thermal bremsstrahlung cooling luminosity $\dot{\varepsilon}_{\rmn{th}} \propto n_\rmn{e} n_\rmn{i}\sqrt{T}$ for cells with $0.2 \, \rmn{keV} < k_\rmn{B}T < 10 \, \rmn{keV}$, where $n_\rmn{e}$ and $n_\rmn{i}$ are the number densities of electrons and ions, respectively. After 1~Gyr, CR protons stabilise at approximately a few percent of the thermal pressure in the central 50~kpc, due to continuous jet activity in this maintenance-phase. Furthermore, in alignment with the results of \citetalias{Ehlert2023} and as shown by the radial profiles, our AGN jet feedback establishes a cycle of self-regulation after ${\sim} 0.5$ Gyr.

\section{FR classification}\label{sec:appendix:FR-classification}

\begin{figure}[ht]
	\centering
	\includegraphics[width=1.\columnwidth]{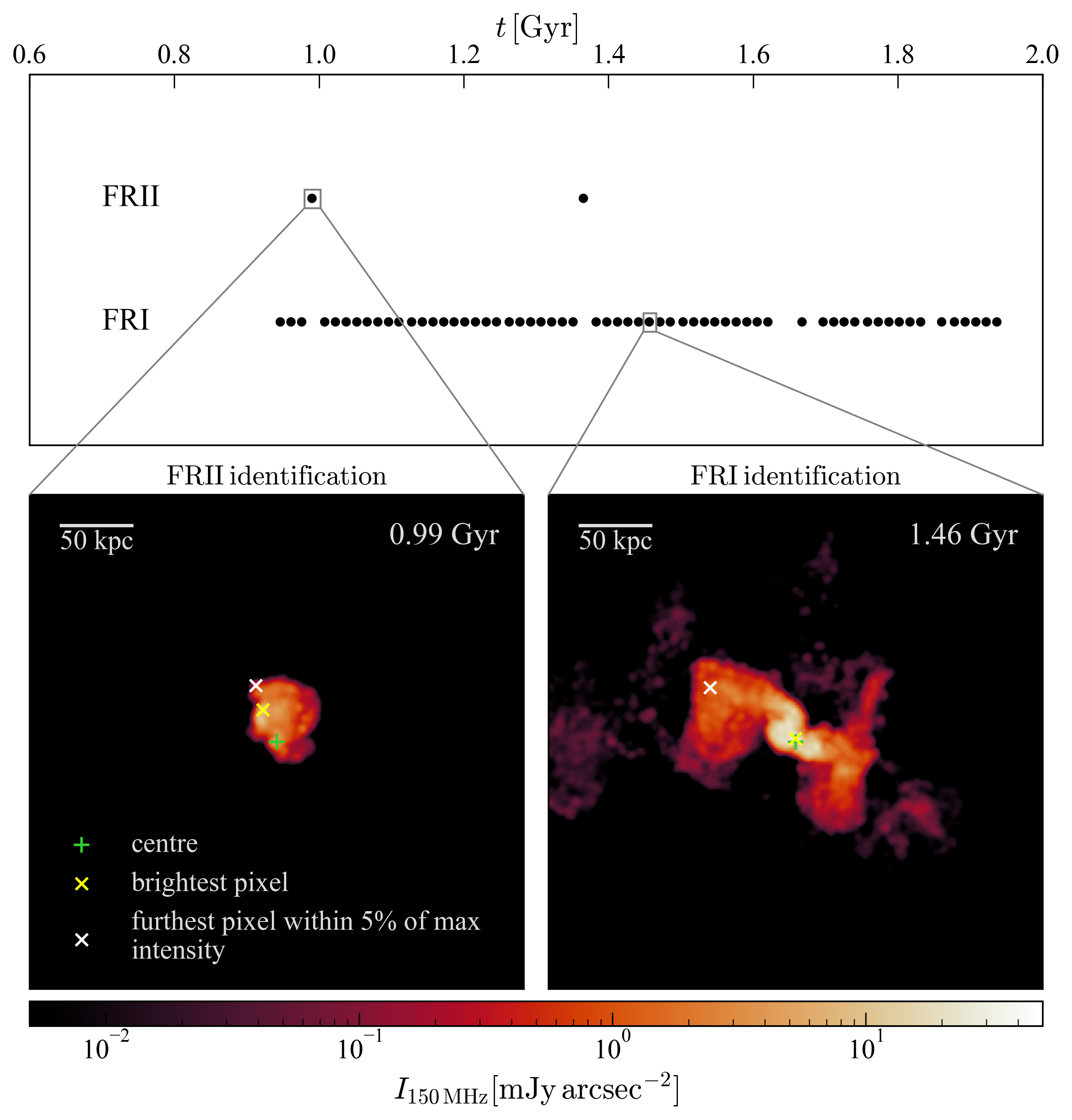}
	\caption{Time evolution of FR classification indices obtained following the original criteria defined in \citet{Fanaroff1974}, calculated for all times where the jet is active in the maintenance phase ($t > 0.9$~Gyr). We obtain FRI indices for almost all times, except at those times where the morphology cannot be described as twin lobes, at which point the FRI-FRII classification should no longer be applied.}
	\label{fig:FR_index}
\end{figure}

The traditional FR classification \citep{Fanaroff1974} consists of determining whether the source's brightest region is closer to the core (i.e. the host) or to the edge of the source, in comparison the mid-point of the source on a given side. In the former, the source is identified as an FRI source, while it is classified as an FRII in the latter. First, we only choose images from our simulations when the jet is active. We use radio intensity maps at 150~MHz to find the mid-point of the source (defined as the pixel corresponding to half of the extent of the source). To find the extent of the source, we use radio intensity maps at 150~MHz, identify all pixels with at least 5$\%$ of the maximum intensity, and choose the one with the largest radial position. We then find the mid-point of the source, defined as the pixel corresponding to half of the extent of the source. Finally, we find the brightest pixel of the source. We then classify our images as FRI or FRII depending on the location of this brightest pixel relative to the mid-point. This procedure is illustrated in Fig.~\ref{fig:FR_index}, where we show the locations of the host in green (in our case, the SMBH), the furthest pixel within 5$\%$ of the maximum radio intensity in white, and the brightest pixel in yellow. 

This yields an FRI identification rate of $\approx 97\%$, as shown in Fig.~\ref{fig:FR_index}. However, upon closer inspection, we found that images classifying FRII radio galaxies did not resemble the traditional morphology expected for such objects. Rather, it reflects the limits of such a classification procedure when dealing with complex morphologies where no clear opposing polar lobes are visible, as shown in Fig.~\ref{fig:radio_maps}. This is also why FR classification procedures usually involve manual verification even when using automated algorithms.

\begin{figure*}[ht]
	\centering
	\includegraphics[width=2.\columnwidth]{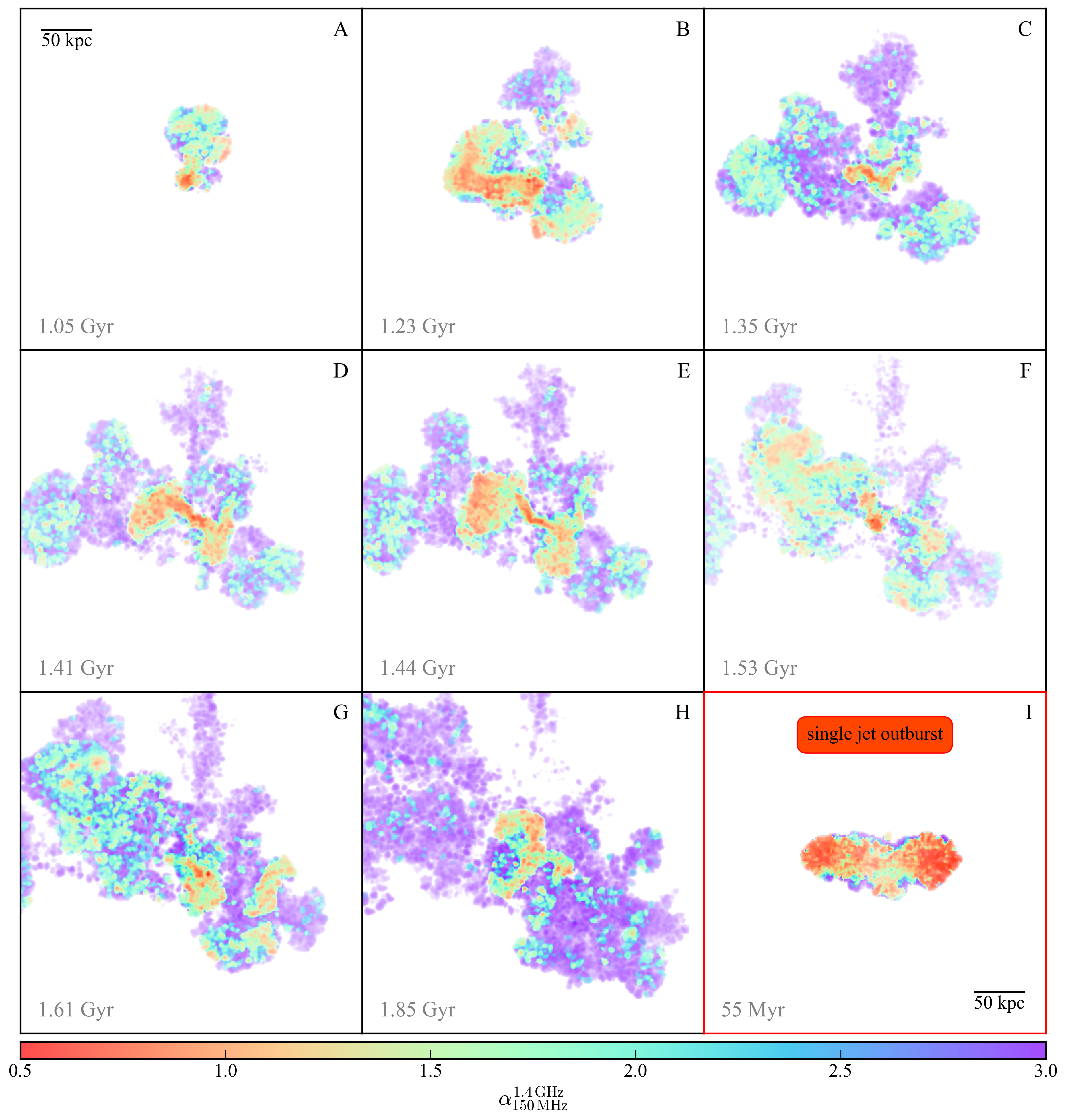}
	\caption{Spectral index maps calculated between 150~MHz and 1.4~GHz, obtained using radio intensity maps smoothed with a 2D Gaussian beam $\rm{FWHM} = 2$~kpc, smaller than beam sizes used throughout the paper in order to highlight small-scale spectral index variations.}
	\label{fig:many_spectral_index_maps}
\end{figure*}

\section{Spectral index maps}\label{sec:appendix:spectral-index-maps}

In Fig.~\ref{fig:many_spectral_index_maps}, we show spectral index maps corresponding to the radio images at 150~MHz presented in Fig.~\ref{fig:radio_maps}. Whereas some of the complex morphologies observed in our radio maps make interpretations on the state of the plasma difficult, these spectral index maps indicate where more recently accelerated plasma lies.

\begin{figure}[ht]
	\centering
	\includegraphics[width=1.\columnwidth]{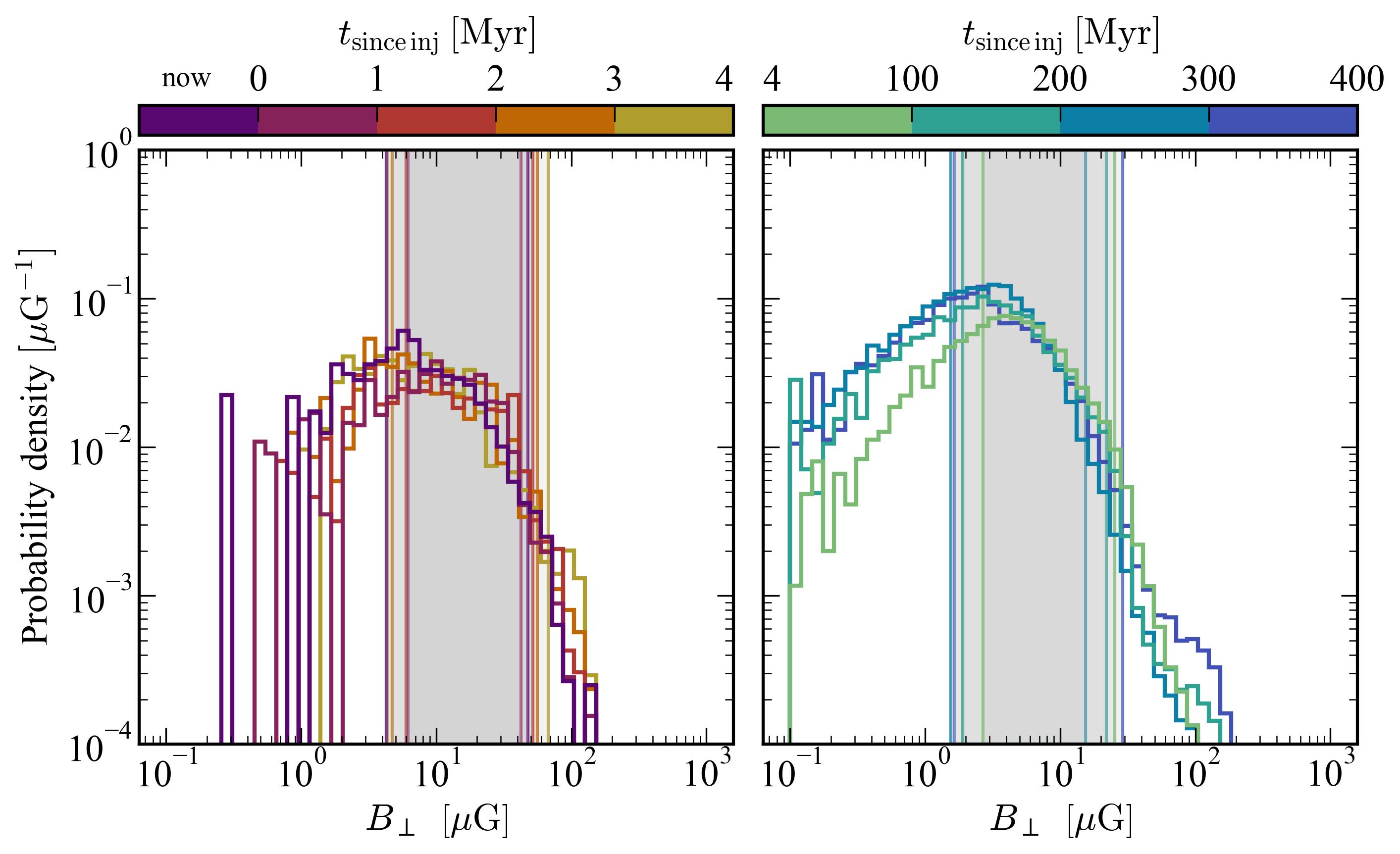}
	\caption{Magnetic field strength histograms for the electron age distributions shown in Fig.~\ref{fig:spectra_binned_by_time_since_inj}. We display the 10th and 90th percentiles of each distribution with gray bands, each delimited by coloured lines matching the corresponding distribution.}
	\label{fig:Bfield_histograms}
\end{figure}

\section{Magnetic field strengths}\label{sec:appendix:Bfields-histograms}

In Fig.~\ref{fig:Bfield_histograms}, we show magnetic field strength histograms, where each line corresponds to a specific electron age distribution (same as shown in Fig.~\ref{fig:spectra_binned_by_time_since_inj}). The magnetic field strengths broadly range from around 0.2 to 100~$\mu \rmn{G}$ with the 10th to 90th percentile values ranging from around 2 to 40~$\mu \rmn{G}$. The magnetic field strength distribution extends to somewhat lower values in the case of older CR electron populations, reaching strengths of 0.1~$\mu \rmn{G}$, although the 10th to 90th percentile range does not decrease significantly.

\end{appendix}

\end{document}